\documentclass[%
 reprint,
 amsmath,amssymb,
 aps,
 prapplied,
 floatfix,
]{revtex4-1}

\usepackage{graphicx}
\usepackage{dcolumn}
\usepackage{bm}
\usepackage{hyperref}
\usepackage[dvipsnames]{xcolor}
\usepackage{tabularx}
\begin{document}

\title{Contact-Governed Macroscopic Signatures of Strain-Induced Valley Sorting in MoS$_2$ Field-Effect Transistors}

\author{Suhas Suresh Bharadwaj}
\email{f20230029@dubai.bits-pilani.ac.in}
\affiliation{Department of Electrical and Electronics Engineering, BITS Pilani, Dubai Campus, Dubai, United Arab Emirates}

\date{\today}

\begin{abstract}
Strain engineering has emerged as a leading route to control valley-selective transport in monolayer transition metal dichalcogenides. But, whether this quantum effect survives inside a fully contacted device remains an open question. Quantum transport methods capture the underlying valley deflection well, but are limited to isolated nanoscale fragments, leaving it unclear whether a complete device would register strain-induced valley sorting in the bulk channel, at the contacts, or barely at all. Here, the quantum valley-deflection tensor of strained monolayer MoS$_2$ is embedded directly into a macroscopic Poisson-drift-diffusion model, extending this physics for the first time from isolated nanoscale fragments to a complete, contact-inclusive device. Applying this framework to a 300~nm monolayer MoS$_2$ transistor with a 3~nm HfO$_2$ gate dielectric at 77~K, under a localized 6.14~T pseudo-magnetic strain field, shows that the resulting 63.6\% suppression of the ON-state current is not a bulk property of the channel at all. It is set almost entirely by the source and drain metal-semiconductor interfaces. The full suppression is reproduced when only the 15 nm contact regions are strained, and nearly vanishes when the channel interior alone is strained, identifying the metal contact, rather than the 2D channel itself, as the dominant site of strain-induced valley physics in a device. However, observing it requires tunneling-transparent, degenerately-doped contacts, since classical Ohmic barriers overwhelm the topological signal by orders of magnitude. A geometry-independent transmission-line analysis reduces this contact-localized behavior to a single valley-topological contact resistance ($R_{c,V}$) and a predictive inverse-length scaling law. The resulting characterization framework should carry over to any strain-engineered 2D-material heterostructure and can be tested directly with standard multi-length contact structures. Taken together, these results identify the metal-semiconductor interface, not the channel, as the decisive location for engineering and detecting valley-selective transport in real 2D-material devices.
\end{abstract}

\maketitle

\section{Introduction}

Valleytronics, which encodes information in the degenerate momentum extrema (valleys) of a crystal band structure, has emerged as one of the most actively pursued directions in two-dimensional materials research \cite{Rycerz2007, Schaibley2016}. Part of this interest is driven by a broader technological pressure, the scaling of conventional silicon-based complementary metal-oxide-semiconductor (CMOS) logic is approaching fundamental physical limits, and as subthreshold leakage currents and power dissipation densities increase at sub-nanometer nodes, alternative state-variables beyond classical electron charge are being sought to sustain computational performance \cite{Theis2010, Nikonov2013}. But valley-selective transport is equally compelling on its own terms, as a distinct quantum degree of freedom native to a rapidly expanding family of 2D materials.

Monolayer transition metal dichalcogenides (TMDs), specifically Molybdenum Disulfide (MoS$_2$), are at the forefront of this paradigm shift. Due to the breaking of spatial inversion symmetry and the presence of strong spin-orbit coupling, the $K$ and $K'$ valleys at the edges of the 2D Brillouin zone become energetically degenerate but quantum-mechanically distinct, governed by valley-contrasting Berry curvature \cite{Xiao2012, Mak2014}. The foundational works by Cao \textit{et al.} \cite{Cao2012} and Zeng \textit{et al.} \cite{Zeng2012} demonstrated that dynamic control over these valleys can be achieved via optical pumping, utilizing circularly polarized light to exploit valley-dependent optical selection rules. Furthermore, intrinsic valley-contrasting Berry curvature can drive anomalous transverse velocities under in-plane electric fields, as demonstrated by purely electrical nonlocal measurements \cite{Xiao2012, Mak2014, Gorbachev2014}. While these mechanisms successfully provided definitive proof of the underlying quantum physics, their reliance on cryogenic optical excitation or pristine unstrained lattices complicates integration into scalable, real-world 2D-material devices.

Consequently, extrinsic mechanical strain engineering has emerged as a practical solid-state mechanism for valley manipulation. The physical deformation of the 2D lattice alters the nearest-neighbor hopping integrals, which introduces a valley-dependent vector potential $\mathbf{A}(\mathbf{r})$ into the Dirac Hamiltonian \cite{Guinea2010, Vozmediano2010}. The spatial gradient of this strain generates a synthetic pseudo-magnetic gauge field, $\mathbf{B}_{pseudo} = \nabla \times \mathbf{A}$, which preserves time-reversal symmetry by acting with opposite polarity on the $K$ and $K'$ valleys \cite{Settnes2016}. Following theoretical proposals for this extrinsic Lorentz-like deflection \cite{Guinea2010}, experimental observations of giant pseudo-magnetic fields exceeding 300~T were confirmed in graphene nanobubbles \cite{Levy2010} and hBN-mediated corrugations \cite{Jiang2017}. While mapping equivalent highly localized pseudo-fields directly in TMDs remains experimentally challenging, Conley \textit{et al.} established that uniaxial strain directly modulates the band structure of MoS$_2$ \cite{Conley2013}, validating the mechanical strain-to-gauge translation for TMD materials.

When integrated into a transport channel, this gauge field subjects the carriers to a valley-dependent Lorentz force, driving the Valley Hall Effect (VHE) without the need for external magnetic fields \cite{Low2014, Beconcini2016}. This mechanism is mechanistically distinct from the valley-dependent longitudinal magnetoresistance predicted for gapped two-dimensional semiconductors under a real, externally applied magnetic field \cite{Sekine2018}, which requires bulk Berry-curvature-driven transport and a genuine time-reversal-symmetry-breaking field. The strain-engineered pseudo-magnetic route pursued here requires no external field at all, and, as established below, is structurally localized to the metallic injection contacts rather than distributed through the bulk channel. However, verifying this polarization using standard electrical measurement techniques remains a critical bottleneck. Because standard macroscopic contacts measure scalar charge rather than quantum phase or valley index, isolating the spatial separation of valleys from background electrical signatures remains challenging \cite{Cazalilla2014}.

Despite the extensive theoretical literature detailing the microscopic quantum mechanics of strain-induced gauge fields, there exists a fundamental methodological gap between microscopic theory and macroscopic device behavior. Current state-of-the-art simulations predominantly rely on Non-Equilibrium Green's Function (NEGF) formalisms and Density Functional Theory (DFT) to model topological transport \cite{CarrilloBastos2014, Rostami2015}. As extensively reviewed by Cresti \textit{et al.} \cite{Cresti2016}, while the Kubo and NEGF formalisms isolate these microscopic quantum deflections accurately, they are computationally prohibitive. These full-phase quantum solvers are typically restricted to analyzing isolated nanoscale segments. They struggle to scale up to complete field-effect devices that feature finite access resistances and capacitive high-$\kappa$ gate dielectrics{; a further, geometry-specific obstacle for the non-uniform strain profile treated here is discussed in Sec.~\ref{sec:theory}}. Beyond this computational limit, restricting simulations to small channel fragments obscures the physical origin of the macroscopic transport behavior. It remains unclear whether the Valley Hall Effect in a fully gated device is bottlenecked by the bulk channel itself or by carrier injection at the metallic contacts. Resolving this spatial dependence is essential to determine whether experimental detection requires engineering the active channel, the contact interfaces, or both. To close this gap and resolve this ambiguity, these quantum scattering mechanisms must be bridged into continuous, classical Poisson-drift-diffusion solvers (TCAD).

The central contribution of this work is to extend strain-induced valley deflection, a quantum-mechanical effect established by prior theory in isolated nanoscale fragments, into a full, contact-inclusive prediction of how this physics manifests inside a real 2D-material device. A spatially discretized pseudo-magnetic gauge field is embedded into a standard Poisson-drift-diffusion solver as a localized mobility array, extending the reach of strain-engineered valley physics to complete device geometries at a fraction of the computational cost of pure quantum methods. The framework is applied to a 300~nm monolayer MoS$_2$ field-effect transistor with a 3~nm HfO$_2$ gate dielectric at 77~K to isolate a distinct macroscopic signature of the Valley Hall Effect. A {63.6\%} suppression of the ON-state transconductance is demonstrated by the results without altering the subthreshold electrostatics, a macroscopic readout termed the Valley Topological Magnetoresistance (VTMR). To resolve and quantify the exact physical origin of this effect, a strain profile sensitivity analysis is evaluated. It is revealed by the analysis that the VTMR signature is entirely governed by the carrier injection interfaces rather than the bulk channel. The full {63.6\%} suppression is reproduced when only the metallic source and drain boundaries are strained, while the effect is reduced to $<$1\% when the strain is confined to the channel interior. This contact-localized origin is further quantitatively formalized through a transmission-line-method extraction, which isolates a single geometry-independent valley-topological contact resistance ($R_{c,V}$) and defines a predictive crossover length ($L^*$) that governs the scaling of the macroscopic VTMR signal. An experimental design rule for detecting valley polarization using standard electrical terminals is, thus, provided by this contact-specific behavior, directly relevant to the broader effort to engineer and characterize strain at 2D-material interfaces. Ultimately, this macroscopic drift-diffusion framework is established as a general tool for predicting how strain-engineered valley physics manifests inside real, contacted 2D-material devices.

\section{Theoretical Framework and Tensor Derivation}
\label{sec:theory}

Bridging microscopic quantum dynamics with macroscopic device physics requires a direct mathematical translation. Traditional transport simulations evaluate electrostatic gating and drift-diffusion independently of the quantum topological phase. To build a self-consistent device simulation, it is necessary to derive a continuous bridge that links the localized strain of the crystal lattice to a scalar modification in the standard transport equations.

\subsection{Strain-Induced Vector Potentials in the Dirac Hamiltonian}

The low-energy electronic structure of monolayer MoS$_2$ is governed by the two-band massive Dirac Hamiltonian situated at the $K$ and $K'$ points of the Brillouin zone. The application of a non-uniform mechanical strain breaks the discrete translational symmetry of the lattice. This physical deformation modifies the nearest-neighbor hopping integrals and manifests mathematically as a synthetic vector potential $\mathbf{A}(\mathbf{r})$ coupled directly to the electron momentum. The strained Hamiltonian is expressed as:

\begin{equation}
\mathcal{H}_{\tau} = v_F [(\tau \hat{p}_x + \tau eA_x)\hat{\sigma}_x + (\hat{p}_y + \tau eA_y)\hat{\sigma}_y] + \frac{\Delta}{2}\hat{\sigma}_z
\label{eq:hamiltonian}
\end{equation}

where $v_F$ is the Fermi velocity, $\tau = \pm 1$ represents the valley index corresponding to the $K$ and $K'$ valleys respectively, $\hat{p}$ represents the momentum operator, $\hat{\sigma}$ are the Pauli matrices acting on the two-band pseudospin basis, and $\Delta$ is the intrinsic bandgap \cite{Xiao2012, Vozmediano2010}. 

The synthetic gauge field $\mathbf{A} = (A_x, A_y)$ is proportional to the localized strain tensor components $\varepsilon_{ij}$ through the dimensionless electron-phonon coupling parameter $\beta$. While originally derived as a Gr\"{u}neisen parameter for graphene's $p_z$ orbital network, specific tight-binding derivations for the Mo $d$-orbital character of TMDs yield an analogous gauge-coupling magnitude bounded near $\beta \approx 2.5$ \cite{Rostami2015, Amorim2016}. For a lattice constant $a$, the vector potential evaluates to:

\begin{equation}
\mathbf{A} = \frac{\hbar \beta}{2 e a} \left( \varepsilon_{xx} - \varepsilon_{yy}, -2\varepsilon_{xy} \right).
\label{eq:vector_potential}
\end{equation}

The spatial gradient of this vector potential generates the pseudo-magnetic field $\mathbf{B}_{pseudo} = \nabla \times \mathbf{A}$. A comprehensive mapping of these strain-induced topological phases in 2D materials demonstrates that because the vector potential couples directly with the valley index $\tau$ in the Hamiltonian formulation, the resulting pseudo-magnetic field acts with strictly opposite polarity on electrons residing in opposite valleys \cite{Amorim2016}. This specific coupling mechanism preserves global time-reversal symmetry across the crystal while breaking it locally within each individual valley \cite{Settnes2016}.

\subsection{Semiclassical Dynamics and Valley Deflection}

To describe carrier transport beyond the static Hamiltonian, the semiclassical equations of motion are constructed for an electron wavepacket centered at wavevector $\mathbf{k}$. In the presence of an external in-plane electric field $\mathbf{E}$ generated by the source and drain bias, along with the internal pseudo-magnetic field $\mathbf{B}_{pseudo}$, the coordinate and momentum dynamics are governed by the coupled equations \cite{Xiao2010}:

\begin{equation}
\hbar \dot{\mathbf{k}} = -e\mathbf{E} - e (\dot{\mathbf{r}} \times \mathbf{B}_{pseudo}),
\label{eq:semiclassical_k}
\end{equation}
\begin{equation}
\dot{\mathbf{r}} = \frac{1}{\hbar} \nabla_{\mathbf{k}} \mathcal{E}(\mathbf{k}) - \dot{\mathbf{k}} \times \mathbf{\Omega}_{\tau}(\mathbf{k}),
\label{eq:semiclassical_r}
\end{equation}

where $\mathcal{E}(\mathbf{k})$ is the unperturbed band dispersion and $\mathbf{\Omega}_{\tau}(\mathbf{k})$ is the valley-contrasting Berry curvature intrinsic to the broken inversion symmetry of the MoS$_2$ monolayer \cite{Mak2014, Low2014}. A Lorentz-like force is induced by the cross product of the carrier velocity with the pseudo-magnetic field, deflecting $K$ and $K'$ electrons toward opposite transverse boundaries of the channel. The mechanical Valley Hall Effect is constituted by this spatial sorting mechanism. 

Traditional quantum scattering formalisms isolate this deflection effect accurately but fail to scale into highly dense device architectures featuring capacitive gate dielectrics and finite access resistances. A complete field-effect transistor simulation requires translating these microscopic trajectory deflections into a macroscopic drift-diffusion framework suitable for continuum solvers. The relative contribution of $\mathbf{\Omega}_{\tau}(\mathbf{k})$ to macroscopic transport is evaluated in the following section.

\subsection{Macroscopic Tensor Derivation}

To bridge this quantum-to-classical divide, the valley-dependent Lorentz scattering is projected onto the longitudinal mobility tensor using the steady-state Boltzmann Transport Equation (BTE) under the relaxation time approximation \cite{Lundstrom2000}.

While the intrinsic Berry curvature $\Omega_\tau(k)$ generates a baseline anomalous velocity \cite{Mak2014}, it can be safely neglected in the strong-inversion drift regime. At a gate overdrive of 9~V across the 3~nm HfO$_2$ dielectric, the inversion charge density reaches $n \approx 3.6 \times 10^{14}$~cm$^{-2}$. At this density, the extrinsic Hall conductivity driven by the 6.14~T pseudo-magnetic field ($\sigma_{xy} \approx \mu_{xy} n e \sim 10^{-1}$~S/sq) exceeds the typical intrinsic Valley Hall conductivity ($\sim 10^{-5}$~S/sq) by orders of magnitude, dominating the transverse deflection. 

By taking the curl of the valley-dependent vector potential, the pseudo-magnetic field evaluates to $\mathbf{B}_{pseudo} = \nabla \times (\tau \mathbf{A}) = \tau B_z \hat{\mathbf{z}}$. For an electron with an effective mass $m^*$ and an isotropic momentum relaxation time $\tau_m$, the steady-state momentum balance equation in the presence of an external electric field $\mathbf{E}$ is expressed as,

\begin{equation}
m^* \frac{d\mathbf{v}}{dt} = -e(\mathbf{E} + \mathbf{v} \times \mathbf{B}_{pseudo}) - \frac{m^* \mathbf{v}}{\tau_m} = 0.
\label{eq:momentum_balance}
\end{equation}

By defining the intrinsic, zero-field acoustic-phonon-limited carrier mobility as $\mu_0 = e\tau_m/m^*$, the balance equation is divided by $m^*$ to isolate the velocity vector $\mathbf{v}$,

\begin{equation}
\mathbf{v} + \mu_0 (\mathbf{v} \times \mathbf{B}_{pseudo}) = -\mu_0 \mathbf{E}
\label{eq:velocity_balance}
\end{equation}

Expanding the cross product in the two-dimensional transport plane ($x,y$) yields a system of coupled linear transport equations for a specific valley $\tau$,

\begin{equation}
\begin{pmatrix} 1 & \tau \mu_0 B_z \\ -\tau \mu_0 B_z & 1 \end{pmatrix} \begin{pmatrix} v_x \\ v_y \end{pmatrix} = -\mu_0 \begin{pmatrix} E_x \\ E_y \end{pmatrix}
\label{eq:coupled_matrix}
\end{equation}

To extract the macroscopic effective mobility tensor $\boldsymbol{\mu}$ required for the continuity equations of the TCAD solver ($\mathbf{v} = -\boldsymbol{\mu} \mathbf{E}$), the localized conductance matrix is inverted,

\begin{equation}
\begin{pmatrix} v_x \\ v_y \end{pmatrix} = \frac{-\mu_0}{1 + (\tau \mu_0 B_z)^2} \begin{pmatrix} 1 & -\tau \mu_0 B_z \\ \tau \mu_0 B_z & 1 \end{pmatrix} \begin{pmatrix} E_x \\ E_y \end{pmatrix}
\label{eq:inverted_matrix}
\end{equation}

This inverted matrix provides the exact analytical bridge between the quantum valley dynamics and the macroscopic device physics. The off-diagonal components ($\mu_{xy} = \mp \tau \mu_0^2 B_z / [1 + (\mu_0 B_z)^2]$) govern the transverse Valley Hall currents. Because $\mu_{xy}$ scales linearly with $\tau$, the $K$ and $K'$ valley currents propagate in opposite directions toward the transverse boundaries \cite{Beconcini2016}. 

Conversely, the forward propagation along the source-drain axis is governed by the longitudinal diagonal component ($\mu_{xx}$). Because the valley index $\tau$ is squared in the denominator ($(\pm 1)^2 = 1$), the longitudinal mobility attenuation is mathematically identical for both valleys,

\begin{equation}
\mu_{mod}(x) = \mu_{xx} = \frac{\mu_0}{1 + (\mu_0 B_z(x))^2}
\label{eq:mobility_tensor}
\end{equation}

This scalar reduction isolates the longitudinal momentum suppression caused by transverse valley deflection. In the bulk of an ideal two-valley system, the oppositely directed $K$ and $K'$ Hall currents cancel, preventing a net transverse Hall field from forming ($E_y \approx 0$). In this scenario, substituting $E_y = 0$ into Eq.~\eqref{eq:inverted_matrix} directly yields the suppressed longitudinal velocity $v_x = -\mu_{mod}E_x$. Furthermore, in practical device geometries, the highly conductive source and drain metallic electrodes explicitly short-circuit any transverse electrostatic fields at the boundaries, enforcing $E_y = 0$ at the injection interfaces.

{This scalar mapping is a phenomenological bridge rather than a first-principles derivation of where the resulting suppression should localize. Eq.~\eqref{eq:mobility_tensor} is motivated by an idealized Hall-bar picture with a genuine in-plane transverse coordinate, but the finite-element implementation of Sec.~\ref{sec:device} never solves a separate transverse dimension, since the simulated geometry is a conventional vertical device cross-section with all resistances normalized per unit device width in the standard TCAD convention. $\mu_{mod}(x)$ is applied directly as a local material parameter along $x$, and the transverse dynamics motivating its functional form are not themselves solved for at the device level. The contrast between contact-proximal and bulk-interior strain in Fig.~\ref{fig:sensitivity} accordingly follows not from any transport boundary condition on $E_y$ or $J_y$, but from mechanisms already established elsewhere in this work, two-dimensional current crowding at the metal overlap (Sec.~\ref{sec:tlm}) \cite{Berger1972, ReevesHarrison1982} in the Ohmic regime and injection-velocity-limited ON-current in the gated regime (Sec.~\ref{sec:gated}) \cite{Lundstrom2000}, where a bottleneck near the source dominates the saturated current far more than one deeper in the channel. A first-principles account would require a valley-resolved, two-population drift-diffusion treatment tracking $n_K(x,y)$ and $n_{K'}(x,y)$ through an intervalley relaxation term, a multi-dimensional extension beyond the single-fluid framework used here.}

{A related concern is whether drift-diffusion itself, which assumes carriers respond instantaneously to the local field, remains adequate at this length scale. The carrier density here places the system deep in the degenerate limit: the exact 2D density-of-states relation (derived below alongside the associated effective-mass correction) gives $E_F-E_C\approx0.96$~eV, roughly $145\,k_BT$ at 77~K, so the appropriate velocity scale for a relaxation length is the Fermi velocity $v_{F,n}=\sqrt{2(E_F-E_C)/m^*}\approx8.7\times10^5$~m/s rather than a thermal velocity. The resulting relaxation length, $v_{F,n}\tau_m\approx660$~nm, exceeds the 15~nm strain feature by more than forty times and the full 300~nm channel by more than two, indicating the concern is not confined to the immediate vicinity of the contacts but plausible throughout the device, placing it in a regime where non-local transport and velocity overshoot \cite{Lundstrom2000} could introduce real quantitative deviation from Eq.~\eqref{eq:mobility_tensor}'s local-equilibrium closure. The drift velocity implied by the applied bias also meets or modestly exceeds $v_{F,n}$ in the Ohmic/TLM baseline ($V_{DS}=1.0$~V, $v_{drift}/v_{F,n}\approx1.16$), though it stays comfortably sub-Fermi in the gated configuration used for the headline suppression ($V_{DS}=0.1$~V, $v_{drift}/v_{F,n}\approx0.12$). None of this affects the comparative conclusions of Sec.~\ref{sec:discussion} (Fig.~\ref{fig:sensitivity}), which compare matched self-consistent drift-diffusion solutions against each other, but it adds quantitative uncertainty to the absolute values of $R_{c,V}$ and the suppression percentages beyond the effective-mass correction below. A fully non-local transport treatment would be needed to quantify this precisely, for instance an ensemble Monte Carlo solution of the Boltzmann transport equation extended with a valley-resolved pseudo-magnetic force term, building on existing open-source semiclassical transport frameworks for monolayer MoS$_2$ \cite{mos2emc2026}, left for future work.}

A direct microscopic (NEGF) benchmark of Eq.~\eqref{eq:mobility_tensor} is left similarly unresolved, and for a reason specific to this system rather than a generic computational cost. Because the pseudo-vector potential of Eq.~\eqref{eq:vector_potential} is site-dependent by construction, tracking the spatial gradient of a non-uniform strain field rather than a constant Peierls phase, it breaks the translational symmetry that ordinarily allows a real-space Green's function treatment to be reduced to a small unit cell propagated via Bloch modes. Every lattice site within the strained region requires an independently Peierls-substituted hopping term over the full 300~nm channel, while the degenerate carrier statistics established above require the valley-resolved contact self-energies to be evaluated over an energy window set by $E_F-E_C$ rather than the few-$k_BT$ window sufficient for a non-degenerate system, self-consistently coupled to the real gate-stack electrostatics of Sec.~\ref{sec:device}. This is precisely the regime identified in the Introduction as intractable for full device geometries \cite{Cresti2016}, and a full treatment is left for future work.

{Independent of these open questions, }while the macroscopic Poisson solver operates on a single carrier fluid, this symmetric mobility reduction {still }reflects the underlying two-valley physics. Unlike a real external magnetic field, which breaks time-reversal symmetry and drives all electrons to one edge (generating a very high net Hall voltage), the pseudo-magnetic gauge field preserves time-reversal symmetry, deflecting $K$ and $K'$ electrons in opposite directions. Because the global transverse Hall voltage cancels out entirely ($\sum_{\tau} E_{y,\tau} = 0$), and because the extraction is exclusive to the scalar longitudinal current ($I_D$), the unified scalar tensor $\mu_{xx}$ accurately captures the symmetric longitudinal momentum choke for both valleys simultaneously without requiring a split two-fluid continuity solver. This 1D spatial mapping captures the localized mobility collapse at the boundaries, allowing the 2D longitudinal Poisson solver to model the boundary-choked current without needing a computationally expensive 3D electrostatic domain \cite{Scharfetter1969}. By evaluating Eq.~\eqref{eq:mobility_tensor} across a highly resolved 1~nm continuous spatial grid, the quantum topological scattering mechanism is embedded directly into the classical Poisson solver. 

For this semiclassical tensor mapping to remain valid, the system must operate within specific physical boundaries. At a targeted cryogenic temperature of 77~K, the intrinsic acoustic phonon-limited mobility of monolayer MoS$_2$, particularly under the strong dielectric screening present at high carrier densities ($n \approx 3.6 \times 10^{14}$~cm$^{-2}$), is evaluated at $\mu_0 \approx 3000$~cm$^2$/Vs based on first-principles deformation potential theory \cite{Kaasbjerg2012}. {At the true peak gauge field of 6.14~T, the corresponding Hall parameter evaluates to $\omega_c \tau_m = \mu_0 B_z = 1.84$. While a Hall parameter exceeding unity typically implies the onset of Landau quantization, at 77~K the thermal energy ($k_B T \approx 6.64$~meV) is roughly a factor of four larger than the pseudo-Landau level spacing at 6.14~T ($\hbar \omega_c \approx 1.57$~meV for $m^* = 0.45 m_0$).} Even at the maximum simulated gauge limit of 7~T, the pseudo-Landau spacing ($\hbar \omega_c \approx 1.79$~meV) remains comfortably obscured by thermal broadening. This energy separation prevents the formation of discrete Landau levels and substantially broadens the quantum oscillations, ensuring the continuous semiclassical Drude-Boltzmann tensor remains a valid approximation \cite{Ashcroft1976}.

{A separate concern is the effective-mass approximation itself at these densities. The exact 2D density-of-states relation, with the correct conduction-band degeneracy ($g=4$, from two nearly spin-degenerate $K$/$K'$ valleys), gives $E_F-E_C\approx0.96$~eV at $n_s=3.6\times10^{14}$~cm$^{-2}$ in the degenerate limit, roughly 53\% of the 1.8~eV bandgap. A Kane-model estimate, $m^*(E)\approx m^*(0)(1+2E/E_g)$, indicates the effective mass at $E_F$ could be enhanced by a factor of order 2, a genuine limitation rather than a negligible correction. This affects the reported numerical values of both the Sec.~\ref{sec:gated} suppression percentages and $R_{c,V}$ itself (Sec.~\ref{sec:tlm}), since both derive from simulations using a fixed, band-edge $\mu_0$, but not the existence of $R_{c,V}$ as a geometry-independent quantity or its inverse-length scaling, which follow from the functional form of Eq.~\eqref{eq:mobility_tensor} rather than the specific value of $m^*$. A full treatment requires an energy-dependent effective mass, left for future work, with reduced gate overdrive offering a direct lever to suppress this correction at the cost of signal magnitude.} Furthermore, in the OFF-state, the longitudinal current is limited by thermionic emission over the channel potential barrier rather than the drift mobility, making the anomalous velocity contributions negligible to the macroscopic subthreshold electrostatics. 

To maintain the required valley coherence, the topological valley polarization must be preserved across the 300~nm channel. Using the MoS$_2$ Fermi velocity $v_F \approx 0.5 \times 10^6$~m/s \cite{Xiao2012}, the intervalley scattering time $\tau_{iv}$ driven by zone-edge optical phonons ($\hbar\omega_K \approx 40$~meV) \cite{Kaasbjerg2012} scales inversely with the phonon occupation number, proportional to $\exp(+\hbar\omega_K / k_B T)$. At 77~K, this low-temperature suppression of phonon scattering yields an estimated intervalley mean free path $\lambda_{iv} > 450$~nm \cite{Kaasbjerg2012}. Because this mean free path exceeds the physical geometry of the device ($\lambda_{iv} > L_{ch}$), valley coherence is sufficiently maintained across the 300~nm channel without rapid intervalley thermalization. Importantly, intervalley scattering in MoS$_2$ requires large-momentum, short-wavelength phonons to allow transitions between the $K$ and $K'$ valleys. Because macroscopic strain gradients are long-wavelength perturbations, they primarily shift the band energies and do not provide the momentum transfer necessary to activate new intervalley scattering channels.

\begin{figure}[htbp]
\centering
\includegraphics[width=1\linewidth]{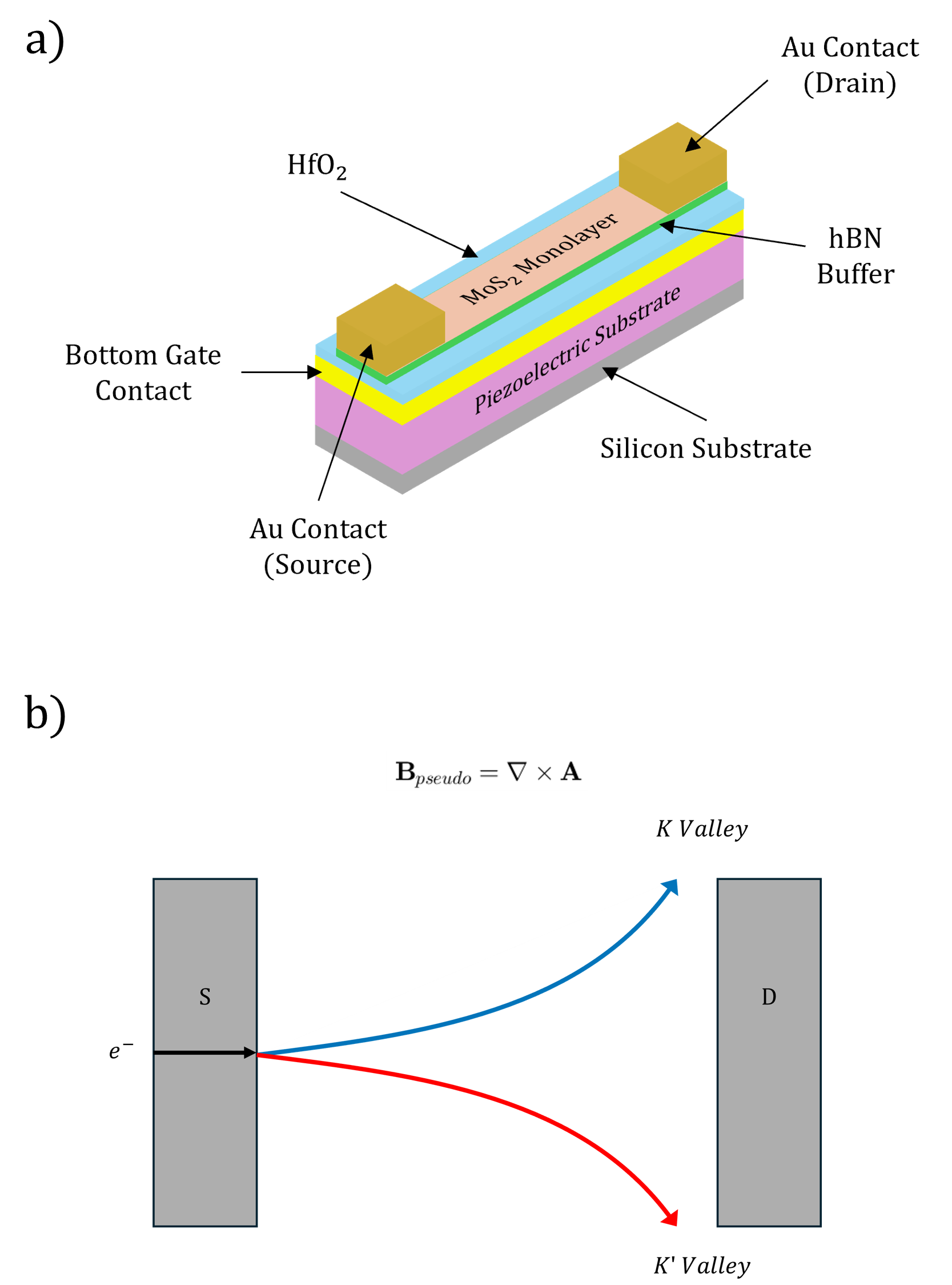}
\caption{Schematic of the high-$\kappa$ monolayer MoS$_2$ Valley-FET architecture. (a) Three-dimensional perspective illustrating the van der Waals heterostructure. From bottom to top, the piezoelectric actuator substrate provides the global strain platform, the back-gate electrode and 3~nm HfO$_2$ gate dielectric form the electrostatic gate stack, and the hBN-buffered MoS$_2$ monolayer channel carries the source and drain contacts. (b) Top-down quantum transport dynamics demonstrating the strain-induced pseudo-magnetic field separating $K$ and $K'$ valley electrons to opposite transverse edges via the Valley Hall Effect.}
\label{fig:concept}
\end{figure}

\section{Device Architecture and Computational Methodology}
\label{sec:device}

Evaluating this analytical tensor framework in a realistic device requires a carefully controlled computational model. The topological transport effects must be isolated from parasitic electrostatic artifacts to ensure that any observed transconductance suppression originates from the pseudo-magnetic gauge field rather than standard interface scattering.

\subsection{Monolayer Channel and Gate Stack Definition}

The simulated architecture is a back-gated field-effect transistor constructed within the Silvaco VictoryDevice finite-element environment. The channel consists of a pristine monolayer of MoS$_2$ with a longitudinal length of 300~nm and an effective out-of-plane electrical thickness of 1~nm, approximating the finite electrostatic thickness of a single physical monolayer. The intrinsic bandgap is set to 1.8~eV with symmetric electron and hole effective masses of $0.45 m_0$, precisely reflecting the massive Dirac model parameters established for the direct-gap monolayer by Xiao \textit{et al.} \cite{Xiao2012}. 

To evaluate the topological transport under realistic field-effect conditions, the MoS$_2$ is buffered by an atomically flat layer of hexagonal boron nitride (hBN), which is then integrated with a 3~nm HfO$_2$ high-$\kappa$ gate dielectric. The hBN layer is essential. Its atomically flat van der Waals surface physically shields the MoS$_2$ channel from remote surface roughness and surface optical (SO) phonons generated by the HfO$_2$ oxide \cite{dean2010boron, Cui2015}, which would otherwise degrade the phonon mobility at 77~K. To preserve the baseline electrostatics and isolate the strain physics, this ultra-thin hBN buffer is computationally implemented as a 1~nm ideal transparent boundary condition in the electrostatic simulation rather than a series capacitor. Consequently, the high dielectric constant of the underlying 3~nm HfO$_2$ ($\kappa \approx 22$) provides the necessary capacitive coupling to drive the channel into deep inversion \cite{Radisavljevic2011}. To eliminate any structural ambiguity in the electrostatic boundary conditions, the physical device stack from bottom to top is defined as follows (Fig.~\ref{fig:concept}): the rigid piezoelectric actuator substrate, which provides the global structural strain-generation platform for the device without participating electrically in the gate stack, the global metallic back-gate electrode deposited directly on this substrate, the 3~nm HfO$_2$ gate dielectric, the 1~nm hBN buffer layer, the MoS$_2$ monolayer channel, and finally the top-deposited metallic source and drain contacts. The gate capacitance entering the inversion-charge calculation of Sec.~II~C is therefore set exclusively by the 3~nm HfO$_2$ layer, with the piezoelectric substrate contributing no series capacitance to the electrostatic gate stack.

\subsection{Finite-Element Discretization of the Pseudo-Magnetic Tensor}

Commercial technology computer-aided design simulators compute macroscopic charge fluxes but do not natively solve for synthetic gauge fields or valley-dependent momentum trajectories. This limitation is overcome by discretizing the continuous spatial mobility tensor derived in Eq.~\eqref{eq:mobility_tensor} into a highly resolved 2D finite-element mesh. 

\begin{figure}[htbp]
\centering
\includegraphics[width=1\linewidth]{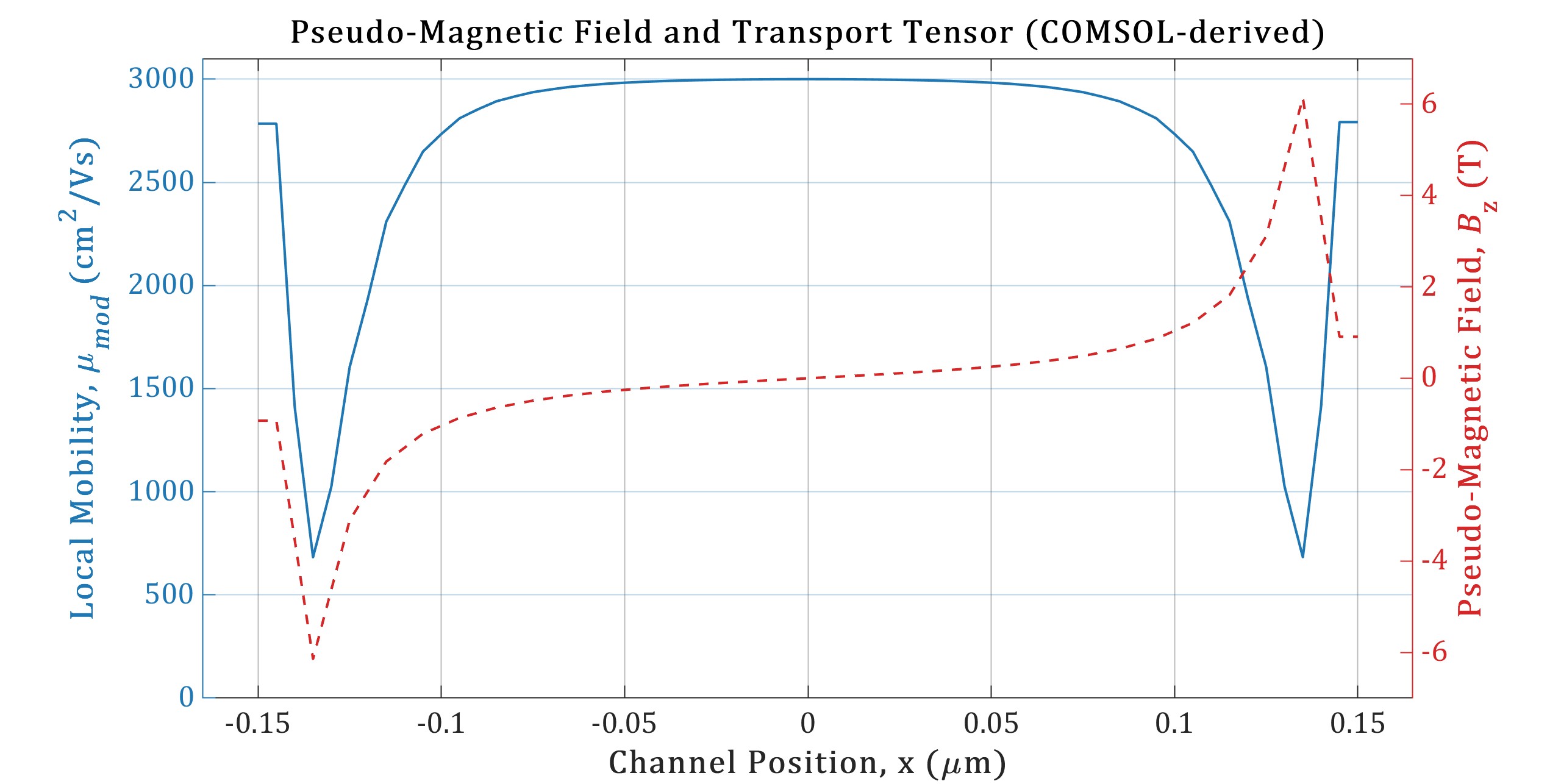}
\caption{Spatial distribution of the normalized pseudo-magnetic gauge field and the corresponding macroscopic transport tensor. {The right axis (red dashed line) traces the finite-element-derived $B_z$ field, peaking at 6.14~T approximately 15~nm inboard of each electrode rather than at the interface itself. The left axis (solid blue line) maps the resulting mobility tensor $\mu_{mod}$, demonstrating the symmetric structural suppression from the intrinsic 3000~cm$^2$/Vs acoustic-phonon limit down to a 698~cm$^2$/Vs minimum, with partial recovery directly at the contact edge.}}
\label{fig:tensor}
\end{figure}

\begin{table*}[htbp]
\centering
\caption{Key physical and simulation parameters used throughout this work.}
\label{tab:parameters}
\begin{tabular}{ll}
\\ \hline\hline
Parameter & Value \\
\hline
Acoustic-phonon-limited mobility, $\mu_0$ & 3000~cm$^2$/Vs \\
Effective mass, $m^*$ (electron, hole) & $0.45\,m_0$ \\
Momentum relaxation time, $\tau_m$ & 0.768~ps \\
Gauge-coupling parameter, $\beta$ & 2.5 \\
Peak pseudo-magnetic field, $B_{max}$ & 6.14~T \\
Peak field location & $\sim$15~nm inboard of contact \\
Minimum contact mobility & 698~cm$^2$/Vs \\
Peak principal strain, $\varepsilon_{max}$ & $\sim$7.0\% \\
Intrinsic bandgap, $E_g$ & 1.8~eV \\
Channel length, $L$ & 300~nm (150--800~nm swept) \\
Contact metal overlap, $L_{contact}$ & 5~nm \\
HfO$_2$ gate dielectric thickness & 3~nm ($\kappa \approx 22$) \\
hBN buffer thickness & 1~nm (transparent BC) \\
Operating temperature, $T$ & 77~K \\
Ohmic sheet density, $n_s$ & $3.5\times10^{14}$~cm$^{-2}$ \\
Gated ON-state carrier density, $n$ & $3.6\times10^{14}$~cm$^{-2}$ \\
Interface trap density, $D_{it}$ (gate oxide) & $3\times10^{12}$~cm$^{-2}$eV$^{-1}$ \\
Contact model (baseline) & ideal Ohmic ($\phi_B = 0$) \\
Richardson constant, $A^{**}$ & 54.08~A/(cm$^2$K$^2$) \\
\hline\hline
\end{tabular}
\end{table*}

{The pseudo-magnetic field is generated by a clamped piezoelectric actuator boundary condition in a finite-element structural mechanics model (COMSOL Multiphysics), rather than assumed in closed form. The strain field is mapped to the vector potential via Eq.~\eqref{eq:vector_potential} and its curl evaluated numerically, yielding a peak field $B_{max}\approx6.14$~T located $\sim$15~nm inboard of each contact rather than at the interface itself, with a local minimum at the edge before decaying into the bulk (Fig.~\ref{fig:tensor}). The associated peak principal strain ($\varepsilon_1$, COMSOL \texttt{solid.ep1}) is $\varepsilon_{max}\approx7.0\%$, within though close to the lower bound of the 6--11\% fracture limit of MoS$_2$ \cite{Bertolazzi2011} (Sec.~\ref{sec:discussion}). {The 15~nm length scale is a prediction of this clamped-boundary finite-element model, not an experimentally measured input. For comparison, measured pseudo-field decay lengths in strained graphene nanobubbles are reported on similar few-to-tens-of-nanometer scales \cite{Levy2010}.}}

The gauge field is mapped statically before running the coupled Newton-Richardson electrostatic iterations \cite{Selberherr1984}. {As mapped in Fig.~\ref{fig:tensor}, the resulting mobility tensor flatlines at the unperturbed limit of 3000~cm$^2$/Vs across the bulk of the channel, dips to a true minimum of 698~cm$^2$/Vs approximately 15~nm inboard of each contact (the closed-form Eq.~\eqref{eq:mobility_tensor} evaluated at $B_z=6.14$~T gives 684~cm$^2$/Vs, since the field peak and the mobility minimum do not exactly coincide in the finite-element solution), and partially recovers to $\sim$2784~cm$^2$/Vs directly at the metal-semiconductor edge itself, reflecting the non-monotonic structure of the finite-element strain solution near the boundary.} To ensure numerical stability across these sharp gradients, the 300~nm MoS$_2$ channel is evaluated via a continuous finite-element spatial discretization. A highly resolved 1~nm meshing grid is utilized along the longitudinal transport axis, providing 15 explicitly defined computational nodes strictly within the 15~nm Gaussian decay region at each boundary. The 270~nm bulk channel is similarly meshed but maintains unperturbed baseline transport. This specific grid resolution was rigorously validated against 0.5~nm mesh control runs to guarantee the convergence of the mobility tensor without numerical discretization artifacts.

\subsection{Numerical Solvers and Transport Physics}

The electrostatic and transport physics are resolved using a coupled Newton-Richardson iteration scheme within Silvaco VictoryDevice. Standard Shockley-Read-Hall (SRH) recombination \cite{Shockley1952, Hall1952} and global Fermi-Dirac statistics are enforced to accurately capture the degenerate electron gas behavior that emerges when the high-$\kappa$ gate drives the monolayer into strong inversion \cite{Taur2013}. To isolate the impact of the strain-induced spatial mobility tensor on macroscopic transport without the confounding influence of high-field effects, the device is deliberately operated in the low-field, linear-response regime throughout this work. All transfer characteristics are extracted at a fixed, small drain-to-source bias ($V_{DS} = 0.1$~V for the gated architecture, and $V_{DS} \leq 1.0$~V for the Ohmic baseline of Sec.~IV\,A), ensuring $V_{DS} \ll V_{GS} - V_T$ across the full gate-voltage sweep. No high-field or velocity-saturation mobility model is invoked. This is a deliberate methodological choice rather than a simplification, it guarantees that any suppression observed in the resulting $I_D$-$V_G$ characteristics originates exclusively from the strain-induced valley-deflection tensor $\mu_{mod}(x)$ of Eq.~\eqref{eq:mobility_tensor}, rather than being convolved with the independent physics of carrier velocity saturation at high longitudinal fields.

The entire computational domain is fixed to a cryogenic ambient temperature of 77~K. This thermal boundary condition is critical for studying topological transport. At room temperature, the transport in MoS$_2$ is heavily dominated by intervalley optical phonon scattering, which degrades the baseline mobility and washes out the delicate synthetic gauge field effects. Freezing out the phonon population at 77~K guarantees that the carrier dynamics are governed almost exclusively by the engineered mechanical strain tensor \cite{Kaasbjerg2012}.

The simulation sequence initializes at a flat-band zero-bias condition before establishing a constant drain-to-source bias. The gate voltage is then swept continuously from the deep subthreshold depletion regime up to a strong inversion state. Both the isolated Ohmic channel transport and the fully gated High-$\kappa$ architecture are simulated to provide a rigorous comparative baseline for the valley topological magnetoresistance extraction.

Table~\ref{tab:parameters} consolidates the key physical and simulation parameters used throughout this work, including several introduced later in the contact- and thermionic-emission analyses of Sec.~\ref{sec:gated} and Sec.~\ref{sec:thermionic_limit}.

\begin{figure*}[t]
\centering
\includegraphics[width=0.85\linewidth]{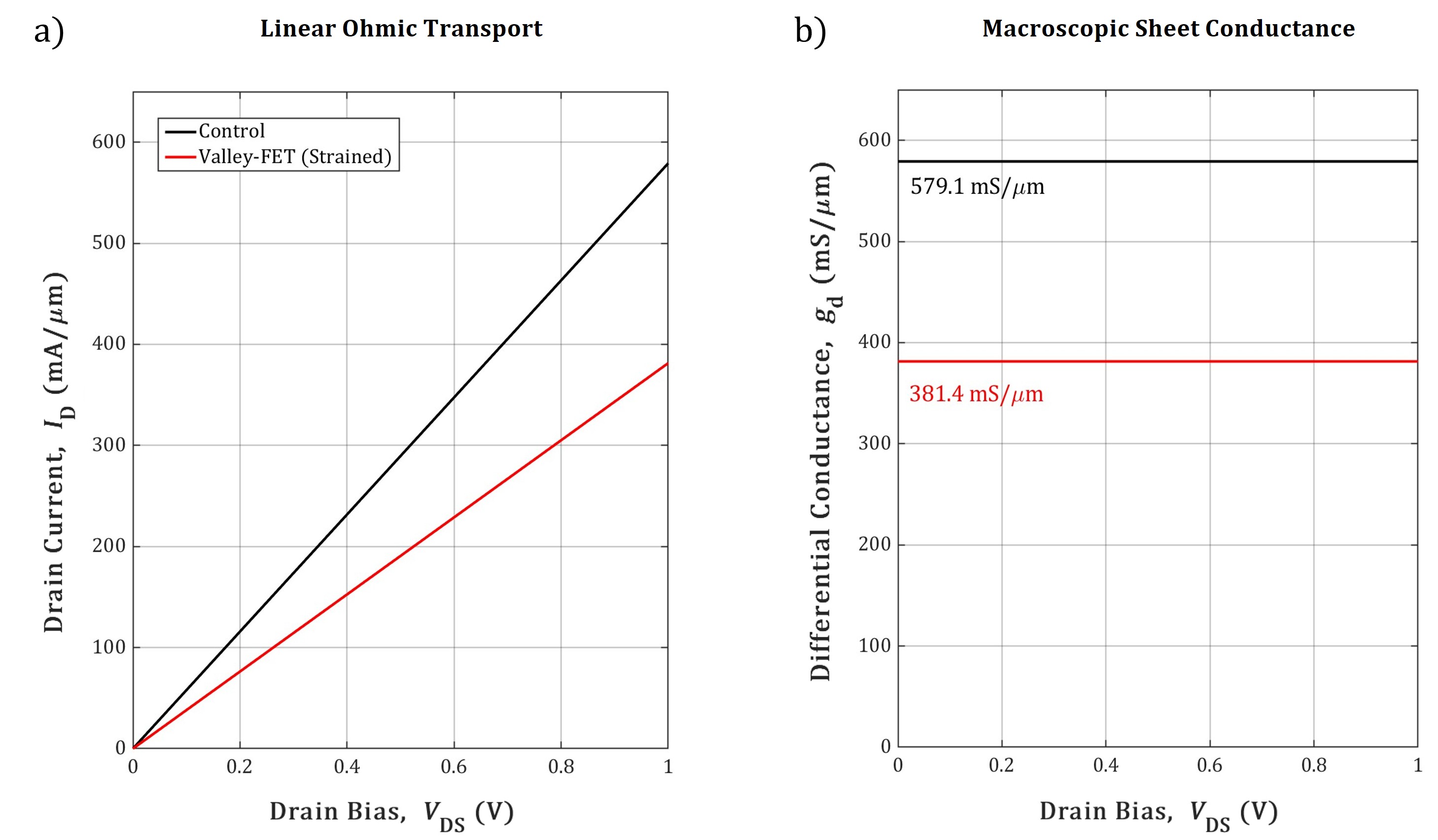}
\caption{Longitudinal macroscopic transport in the idealized Ohmic regime at 77~K. (a) The $I_D - V_G$ transconductance curves demonstrating pure Drude-like linearity in the absence of an oxide barrier. (b) Macroscopic differential conductance ($g_d$) extracted via linear regression confirming that the spatially varying topological strain induces a continuous suppression of the intrinsic sheet conductance from 579~mS/$\mu$m to 381~mS/$\mu$m.}
\label{fig:ohmic}
\end{figure*}

\section{Results and Discussion}

\subsection{Intrinsic Topological Transport in the Ohmic Baseline}
\label{sec:ohmic}

To establish a baseline for the topological momentum suppression, the MoS$_2$ channel is first examined without the high-$\kappa$ gate dielectric. To evaluate the intrinsic sheet conductance at high carrier densities independently of electrostatic gating, the Ohmic channel is simulated with a fixed electrostatic boundary condition, a uniform sheet charge density of $N_D \approx 3.5 \times 10^{14}$~cm$^{-2}$. This establishes the high carrier density Ohmic baseline without relying on unphysical volumetric substitutional impurities. Under these boundary conditions, the channel acts as an Ohmic sheet resistor.

The longitudinal transport characteristics of the bare channel at 77~K are illustrated in Fig.~\ref{fig:ohmic}. Linear current-voltage relationships are exhibited by both the control device and the strained Valley-FET. To maintain the cryogenic thermal equilibrium assumption and prevent high-field Joule heating, the longitudinal bias is fixed at $V_{DS} = 1.0$~V, identical to the bias condition used throughout the channel-length extraction of Sec.~\ref{sec:tlm}. At this bias, the control device yields a steady-state current of 579~mA/$\mu$m. This current is reduced to 381~mA/$\mu$m upon the application of the localized 6.14~T pseudo-magnetic gauge field, corresponding to a Valley Topological Magnetoresistance (VTMR) ratio of exactly 34.1\%.

The macroscopic differential conductance panel in Fig.~\ref{fig:ohmic} confirms that the topological strain induces a continuous step-down in the overall quantum sheet conductance from 579~mS/$\mu$m to 381~mS/$\mu$m (a matching $\sim$34.1\% suppression), in exact quantitative agreement with the independent transmission-line extraction reported in Table~\ref{tab:tlm}. The $K$ and $K'$ electrons are driven to opposite edges of the channel where their forward propagation is throttled by boundary accumulation.

\begin{table}[htbp]
\centering
\caption{Channel-length sweep isolating the geometry-independent valley-topological contact resistance. All resistances are reported per unit device width. \\}
\label{tab:tlm}
\begin{tabular}{cccc}
\hline\hline
$L$ (nm) & $R_{control}$ ($\Omega\cdot\mu$m) & $R_{strained}$ ($\Omega\cdot\mu$m) & VTMR($L$) \\
\hline
150 & 0.8351 & 1.7303 & 51.7\% \\
300 & 1.7267 & 2.6219 & 34.1\% \\
500 & 2.9155 & 3.8108 & 23.5\% \\
800 & 4.6988 & 5.5940 & 16.0\% \\
\hline\hline
\end{tabular}
\end{table}

\subsection{Channel-Length Scaling and the Topological Contact Resistance}
\label{sec:tlm}

The 34.1\% Ohmic suppression reported above is a single-geometry measurement, and it cannot alone distinguish a genuine localized boundary effect from an artifact specific to the 300~nm device studied. To isolate the injection-boundary contribution as a geometry-independent physical quantity, a transmission-line-method (TLM) style channel-length sweep is performed here, following the classical framework developed for characterizing metal-semiconductor Ohmic contacts \cite{Berger1972, ReevesHarrison1982}.

In steady-state Ohmic transport, current continuity forces the terminal current $I$ to remain constant along the transport axis $x$, so the total two-terminal resistance per unit device width follows directly from a series integration of the local resistivity,
\begin{equation}
R_{tot}(L) = \int_0^L \frac{dx}{\sigma(x)} = \rho_s (L - 2\lambda) + 2\int_0^\lambda \rho(x)\, dx,
\label{eq:tlm_integral}
\end{equation}
where $\rho_s = 1/(qn_s\mu_0)$ is the unperturbed sheet resistivity and $\lambda$ is the full spatial extent of the contact-affected region established by the finite-element strain profile of Sec.~\ref{sec:device}. Because the mobility tensor $\mu_{mod}(x)$ is identical at every channel length by construction, the excess resistivity accumulated across each strained boundary defines a single length-independent quantity,
\begin{equation}
R_{c,V} \equiv \int_0^\lambda \big[\rho(x) - \rho_s\big]\, dx,
\label{eq:rcv_def}
\end{equation}
termed here the valley-topological contact resistance. Eq.~\ref{eq:tlm_integral} reduces to the familiar linear TLM form $R_{tot}(L) = \rho_s L + 2R_{c,V}$, so $R_{c,V}$ can be extracted directly from the intercept of a $R_{tot}$ versus $L$ regression, in exactly the same manner specific contact resistance is conventionally extracted from real metal-semiconductor test structures \cite{Berger1972, ReevesHarrison1982}.

To extract $R_{c,V}$, the Ohmic geometry of Sec.~III is re-simulated at four channel lengths (150, 300, 500, and 800~nm) under the identical fixed sheet density $n_s = 3.5\times10^{14}$~cm$^{-2}$ used in Fig.~\ref{fig:ohmic}, with the drain bias fixed at $V_{DS} = 1.0$~V to remain strictly within the linear transport regime. Both the unstrained control geometry and the topologically strained geometry are evaluated at each length using the identical finite-element mesh and Newton solver settings described in Sec.~III.

Table~\ref{tab:tlm} reports the extracted resistances and the resulting length-dependent VTMR ratio at each channel length. A linear regression of the unstrained control data yields a sheet resistivity of $\rho_s = 5.944$~$\Omega$/sq, in close agreement with the independently expected value $\rho_s = 1/(qn_s\mu_0) = 5.945$~$\Omega$/sq computed directly from the acoustic-phonon-limited mobility of Sec.~II, confirming that the finite-element extraction correctly recovers the intrinsic bulk transport parameter with no free fitting. The small negative control intercept, $-0.057$~$\Omega\cdot\mu$m, corresponds to an effective contact transfer-length correction of $\delta L \approx 9.5$~nm, consistent with the 5~nm source and drain metal overlap defined in the device geometry \cite{Berger1972}. A supplementary control run at 650~nm, extrapolated blind from the 150 and 800~nm points to 13 significant digits, confirms this linearity holds across the full simulated range and is not an artifact of any two-point fit.

The strained geometry shares the identical slope $\rho_s$, as required physically since the bulk mobility is untouched by contact-localized strain, but is offset by an additive intercept of $2R_{c,V} = 0.895$~$\Omega\cdot\mu$m, yielding,
\begin{equation}
R_{c,V} = 0.448~\Omega\cdot\mu\text{m}
\end{equation}
per contact. Because $\rho_s$ scales the total resistance linearly with $L$ while $R_{c,V}$ does not, the VTMR ratio itself becomes an explicit function of device geometry,
\begin{equation}
\text{VTMR}(L) = 1 - \frac{\rho_s L + b_0}{\rho_s L + b_0 + 2R_{c,V}},
\label{eq:vtmr_L}
\end{equation}
where $b_0$ is the small control-line intercept associated with the metal overlap correction above. Eq.~\ref{eq:vtmr_L} defines a natural crossover length,
\begin{equation}
L^\ast = \frac{2R_{c,V}}{\rho_s} \approx 151~\text{nm},
\end{equation}
at which the contact and bulk contributions to the total channel resistance become equal. For $L \ll L^\ast$ the device operates in a contact-dominated regime and VTMR saturates toward its maximum value, while for $L \gg L^\ast$ the topological signal is progressively diluted by the unperturbed bulk channel, following the asymptotic $\text{VTMR}(L) \approx 2R_{c,V}/(\rho_s L)$ scaling visible in the rightmost columns of Table~\ref{tab:tlm}. This inverse-length scaling is directly testable and provides a concrete design rule. Shorter-channel Ohmic test structures more efficiently isolate the topological contact signature from the diluting bulk background.

\begin{figure*}[t]
\centering
\includegraphics[width=\textwidth]{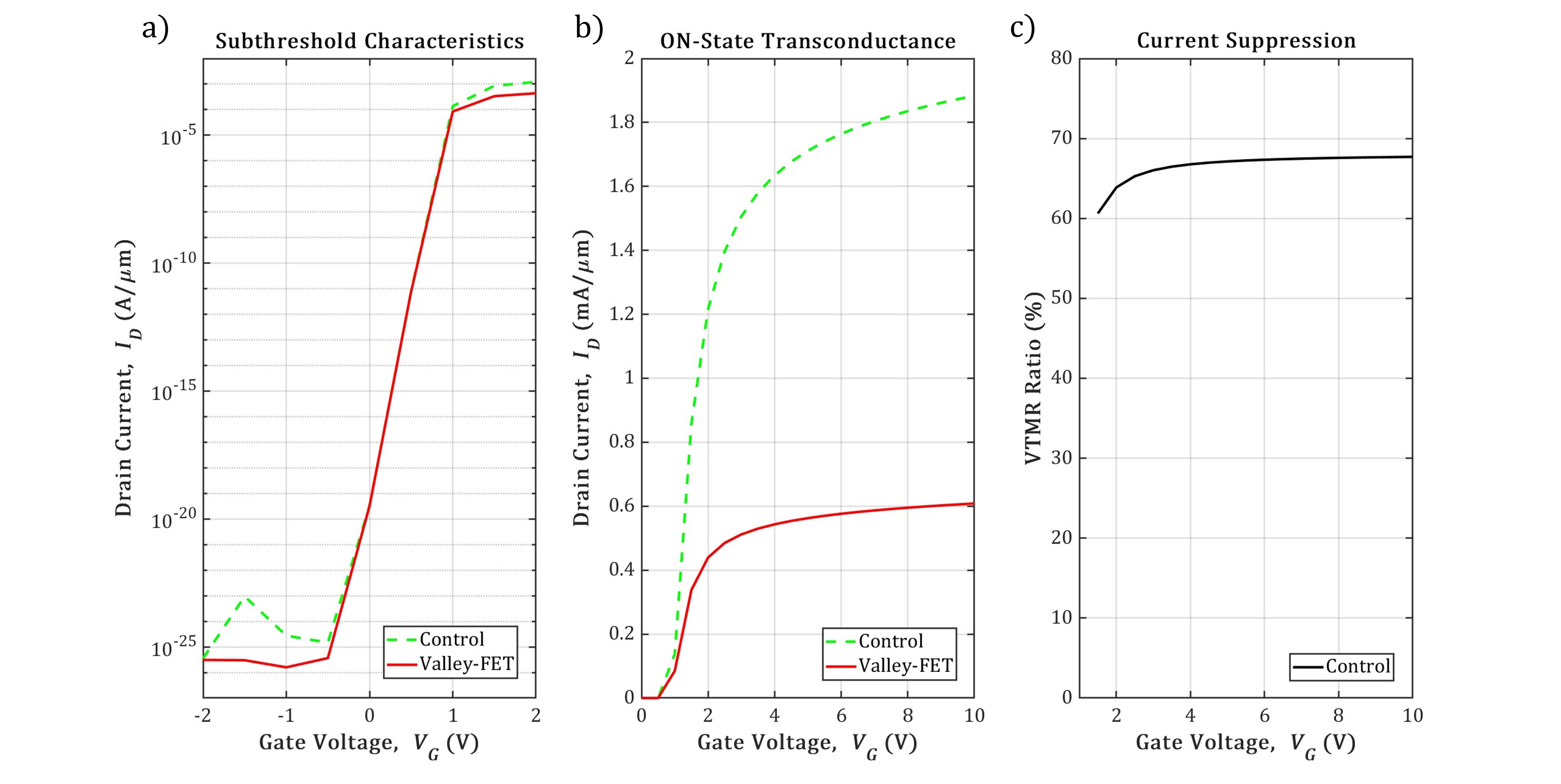}
\caption{Steady-state transport characteristics of the fully gated High-$\kappa$ MoS$_2$ Valley-FET at 77~K. (a) Subthreshold electrostatics demonstrating consistent overlap and a similar subthreshold swing ($\sim 102$ mV/dec), indicating the strain tensor preserves gate integrity and OFF-state limits. (b) Linear ON-state transconductance revealing the macroscopic divergence between the control channel and the topological channel under high carrier density. (c) The Macroscopic Current Suppression mapping the percentage drop, which stabilizes into a {63.6\%} plateau as the transistor enters the deep inversion drift regime.}
\label{fig:highk}
\end{figure*}

\begin{figure*}[t]
\centering
\includegraphics[width=\linewidth]{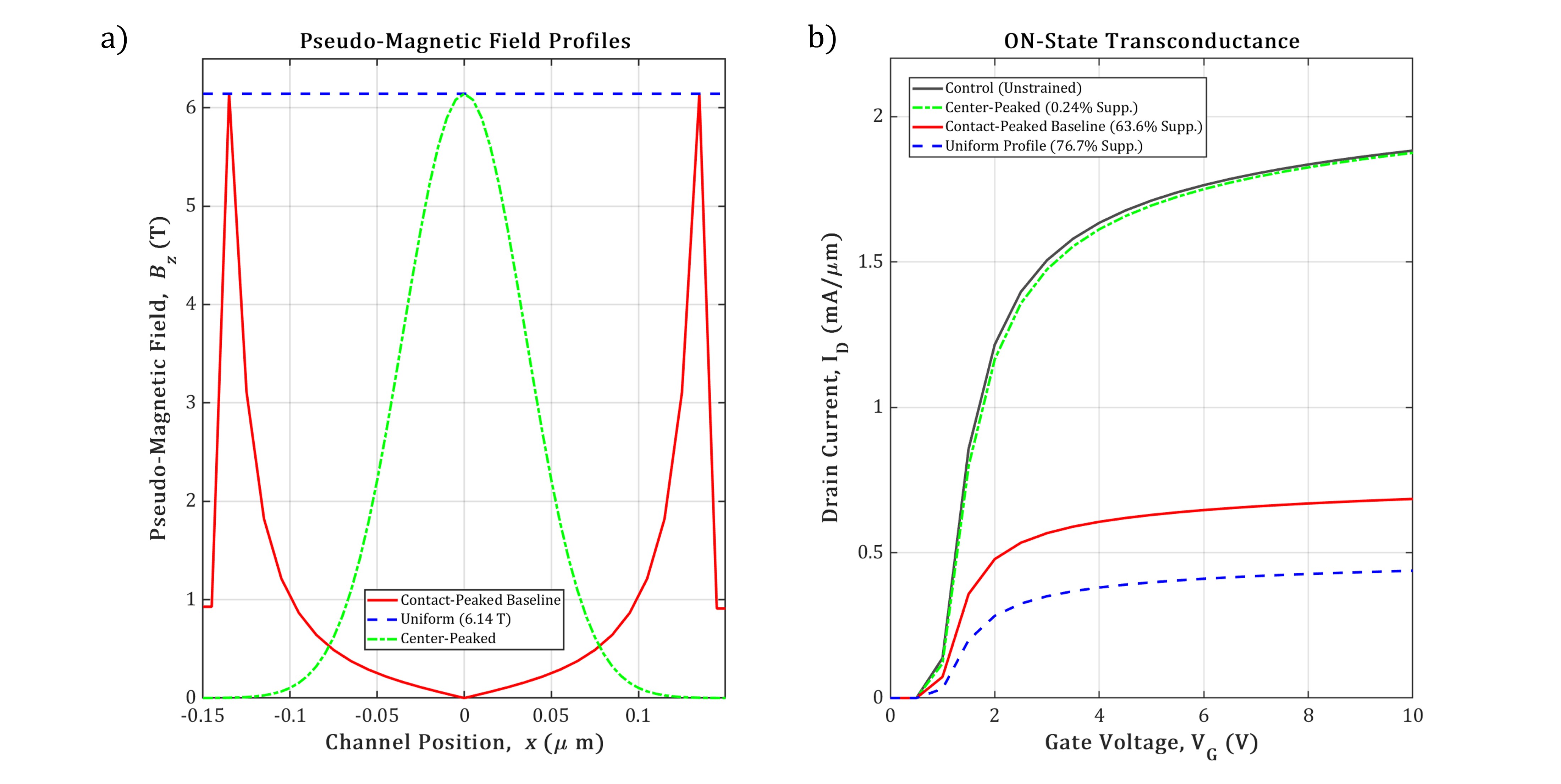} 
\caption{Strain profile sensitivity analysis demonstrating the structural dominance of the source injection bottleneck at 77~K. (a) Spatial distributions of the pseudo-magnetic gauge field $B_z$ across the 300~nm channel. The physically mandated contact-peaked Gaussian profile is compared against two extreme theoretical bounding conditions, a uniform 6.14~T field and a center-peaked field. (b) The corresponding macroscopic transfer characteristics. {The uniform profile yields a larger ON-state suppression than the contact-peaked baseline (76.7\% vs. 63.6\%), indicating that bulk-channel strain contributes measurably to the total suppression rather than adding only negligible series resistance.} Conversely, the center-peaked profile, which leaves the metallic injection boundaries unstrained, exhibits almost zero topological suppression ($<$1\%), confirming that the macroscopic Valley Hall signature is structurally governed by the contact interfaces.}
\label{fig:sensitivity}
\end{figure*}

{The interface trap density $D_{it}$ discussed in Sec.~\ref{sec:gated} is specific to the MoS$_2$/HfO$_2$ gate-oxide interface and does not enter the $R_{c,V}$ extraction above, since the Ohmic geometry used here contains no gate oxide and is evaluated as a fixed-density resistor. $R_{c,V}$ is therefore structurally insensitive to gate-oxide trap density by construction rather than by an argument requiring screening. A distinct, unmodeled question is whether disorder localized at the metal-semiconductor contact itself, rather than the gate oxide, could perturb $R_{c,V}$. The contact region is treated as degenerately doped by charge transfer from the metal (Sec.~\ref{sec:discussion}), and the same Thomas-Fermi screening argument invoked for the gate oxide ($\lambda_{TF}\sim1$--2~nm, well below typical trap spacing) would be expected to similarly screen such disorder, though this has not been directly simulated and is presented as a physically motivated extrapolation rather than a demonstrated result.}

\subsection{Contact-Governed Transport and the Injection Bottleneck}
\label{sec:gated}

Having isolated $R_{c,V}$ as a fixed, geometry-independent penalty attached to the injection boundary in the linear regime, attention now turns to how this same localized contact physics manifests once the device is driven into strong inversion. The integration of the 3~nm HfO$_2$ dielectric alters the boundary conditions of the system, and Fig.~\ref{fig:highk} details the transport characteristics of the fully gated 2D-material architecture.

The logarithmic subthreshold transfer curves in Fig. \ref{fig:highk}(a) show close electrostatic agreement between the strained and unstrained devices. Both architectures exhibit a threshold voltage near 1.0~V and maintain a subthreshold swing of roughly 102~mV/dec. (Note that the jagged plateau observed in deep depletion below $10^{-20}$~A/$\mu$m represents the double-precision numerical noise floor of the finite-element solver rather than physical thermionic leakage.) This broadening from the 15~mV/dec theoretical cryogenic limit results from including a uniform interface trap density ($D_{it} = 3 \times 10^{12}$~cm$^{-2}$eV$^{-1}$) at the MoS$_2$/HfO$_2$ boundary. While this trap density was imported from established SiO$_2$ characterizations \cite{Ghatak2011}, it is utilized here as a conservative upper-bound to rigorously stress-test the topological signature against poor interfacial conditions. While Coulomb scattering from these traps dictates the subthreshold electrostatics, as the device enters strong inversion ($n \approx 3.6 \times 10^{14}$~cm$^{-2}$), the high carrier density provides substantial Thomas-Fermi dielectric screening. At this inversion density, the Thomas-Fermi screening length ($\lambda_{TF} \sim 1-2$~nm \cite{ando1982electronic}) falls well below the mean trap spacing. This effectively neutralizes individual Coulomb scatterers, allowing the acoustic phonon-limited mobility of 3000~cm$^2$/Vs to correctly dominate the ON-state transport.

The macroscopic signature of the Valley Hall Effect is observed exclusively in the strong-inversion drift regime. As shown in the linear transconductance plot, significant divergence between the curves is observed once the device is fully ON. At a gate bias of 10~V, an ON-current of 1.88~mA/$\mu$m is achieved by the control FET. The strained Valley-FET current, however, is substantially reduced to {0.685~mA/$\mu$m}. A {63.6\%} suppression of the ON-state current is yielded by this reduction. 

The amplification of the VTMR from 34.1\% in the bare Ohmic channel to 63.6\% in the gated architecture is a genuine, reproducible feature of the coupled solve, verified against the transfer characteristics of both the contact-peaked and uniform strain profiles. Part of it follows directly from the geometric dilution already derived in Sec.~\ref{sec:tlm}, since at $L=300$~nm the device sits at $L \approx 2L^\ast$, still within the regime where the fixed contact penalty $R_{c,V}$ is diluted by the series bulk resistance (Eq.~\eqref{eq:vtmr_L}).

This dilution alone does not account for the full amplification, since the two architectures are evaluated at nearly identical carrier densities ($n_s \approx 3.5\times10^{14}$~cm$^{-2}$ Ohmic versus $n \approx 3.6\times10^{14}$~cm$^{-2}$ at the ON-state overdrive), ruling out a difference in bulk carrier density as the origin of the residual gap. The residual amplification is instead attributed to the differing electrostatic topology of the two architectures. The bare Ohmic channel is a uniformly doped resistor with no gate coupling, while the gated architecture couples the same mobility tensor $\mu_{mod}(x)$ to a threshold-forming inversion layer beneath a gate electrode spanning the strained boundary, bottlenecking the ON-current through gate-controlled injection velocity at the contact (Sec.~\ref{sec:discussion}) rather than a fixed series resistance alone. This is the most physically plausible explanation consistent with the verified transport data, not an independently decomposed result, and isolating the precise electrostatic contribution would require a spatial carrier-density extraction beyond the scope of this work. Formalizing this gate-coupling mechanism into a closed-form $\text{VTMR}(L)$-type prediction analogous to Eq.~\eqref{eq:vtmr_L} is a natural extension left for future work.

\begin{figure*}[htbp]
\centering
\includegraphics[width=0.85\linewidth]{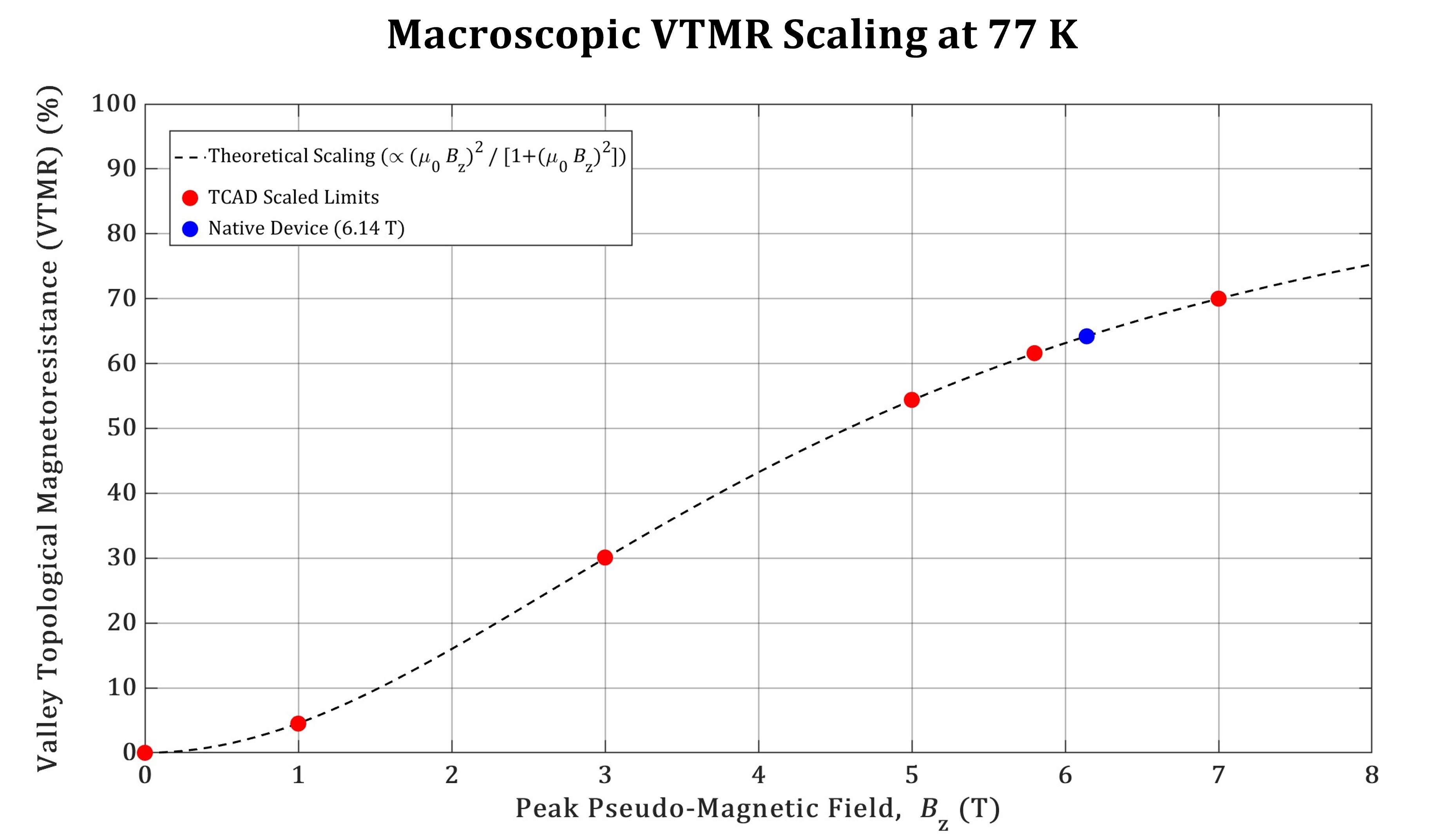}
\caption{Non-linear scaling law of the macroscopic Valley Topological Magnetoresistance extracted at deep inversion ($V_G = 10$~V). The current suppression ratio exhibits a strictly monotonic increase with the peak pseudo-magnetic field magnitude $B_z$, reflecting the quadratic velocity dependence of the transverse valley deflection tensor.}
\label{fig:scaling}
\end{figure*}

To definitively prove that this suppression is driven by the injection bottleneck, a strain profile sensitivity analysis was conducted (see Fig.~\ref{fig:sensitivity}). As illustrated in Fig.~\ref{fig:sensitivity}(a), the physically mandated contact-peaked Gaussian profile was compared against two extreme theoretical bounding conditions, a uniform 6.14~T profile across the entire channel, and a center-peaked profile where the 6.14~T field is isolated in the middle of the channel while the contacts remain unstrained. 

As shown in the macroscopic transfer characteristics in Fig.~\ref{fig:sensitivity}(b), a {76.7\% suppression (0.438~mA/$\mu$m)} is yielded by the uniform profile, {which exceeds the contact-peaked suppression, indicating that bulk-channel strain contributes measurably to the total suppression rather than adding only negligible series resistance}. Conversely, the VTMR is drastically reduced back down to almost zero ({0.24\%}, yielding 1.88~mA/$\mu$m) by the center-peaked profile. It is explicitly confirmed by this 1D spatial extraction that the macroscopic valley-polarization behavior is intrinsically linked to the mobility collapse at the source injection boundary. The field-effect response is, therefore, completely lost if the injection interfaces are left unstrained.

It must be emphasized that this {63.6\%} longitudinal suppression is orders of magnitude larger than typical transverse non-local Valley Hall voltages reported in the literature \cite{Mak2014, Gorbachev2014}. This massive amplification is the direct result of translating a transverse deflection into a longitudinal injection bottleneck. Non-local measurements rely on the secondary diffusion of valley-polarized edge currents, which suffer from rapid intervalley scattering and signal decay. In contrast, this VTMR architecture places the topological scatterer directly at the primary current-choke point of the field-effect transistor, the source injection barrier. By throttling the primary longitudinal drift velocity $v_x$ of Eq.~\eqref{eq:mobility_tensor} directly, rather than measuring a secondary edge diffusion, the device achieves a macroscopically dominant {63.6\%} topological signal.

To demonstrate the physical robustness of the macroscopic suppression model, the scaling behavior of the transconductance is evaluated against varying topological limits. By extracting the steady-state ON-current at a fixed gate bias of 10~V across peak pseudo-magnetic gauge fields of 1~T, 3~T, 5~T, and 7~T, the continuous response of the macroscopic Valley Hall Effect is mapped.

Fig.~\ref{fig:scaling} illustrates the extracted Valley Topological Magnetoresistance ratio as a function of the peak gauge field $B_z$. {The suppression ratio exhibits a distinct non-linear, monotonic dependence, rising from 4.5\% at 1~T to 30.1\% at 3~T, 54.4\% at 5~T, and 70.0\% at 7~T, with the native device operating point falling at 6.14~T and 64.2\%, close to but not exactly the 63.6\% reported elsewhere due to the scaled-parameter extraction used for this sweep. This continuous scaling is consistent with} the saturating functional form derived from the longitudinal mobility tensor in Eq.~\eqref{eq:mobility_tensor}, where the suppression ratio scales proportionally to $(\mu_0 B_z)^2 / [1 + (\mu_0 B_z)^2]${, though the ON-state suppression is systematically below the local-peak-mobility prediction of this formula at every field value, consistent with the injection-velocity-limited (rather than purely local-resistivity-limited) origin of the ON-state suppression discussed above.} It explicitly exhibits quadratic behavior only in the low-field limit ($\mu_0 B_z \ll 1$) before asymptotically flattening at higher fields, validating the computational stability of the spatially discretized finite-element array.

\subsection{Experimental Feasibility and Practical Device Limits}
\label{sec:discussion}

Translating these theoretical projections into a physical device requires navigating several practical engineering constraints. Notably, the transmission-line framework of Sec.~\ref{sec:tlm} is not merely a numerical convenience for this simulation study. Because $R_{c,V}$ is defined identically to a conventional specific contact resistance, it is directly extractable from real fabricated devices using the same multi-length TLM test structures routinely employed to characterize ordinary Ohmic contacts \cite{Berger1972, ReevesHarrison1982}, offering a concrete and immediately actionable experimental protocol for validating this framework independent of the cryogenic gated-transistor measurement discussed below. The simulation isolates the momentum-space vector potential derived from the lattice deformation, while physical strain also generates a scalar deformation potential that modulates the bandgap $E_g$ \cite{Peelaers2012} and, above roughly 1\% uniaxial strain, drives a direct-to-indirect crossover \cite{Conley2013}. Given the 1.8~eV intrinsic gap and the strong electrostatic control of the HfO$_2$ dielectric, this shift barely affects the threshold voltage, and {the required 7.0\%} peak strain confined to the 15~nm injection boundaries places those edges well into this indirect regime. This is inconsequential for the transport bottleneck, however, since the contact region is degenerately doped by charge transfer from the metal, so transport there is governed by a highly occupied, quasi-metallic density of states rather than isolated Dirac cones, and the bottleneck is set by scattering as carriers inject into the pristine 270~nm bulk channel regardless of the local band structure at the extreme edge.

Recent experimental advancements integrating 2D materials with active piezotronics demonstrate that utilizing a clamped PZT substrate to induce stable, localized strain profiles well beyond 1.5\% is mechanically feasible without triggering lattice tearing or oxide delamination \cite{Yang2021}. Experimental demonstrations of piezoelectric substrate actuation in monolayer TMDs (such as WSe$_2$, providing strong cross-material precedent) \cite{schmidt2016reversible} and suspended clamped-membrane MoS$_2$ geometries \cite{lloyd2016band, castellanos2013local} directly validate the localized architecture assumed in this framework.

Quantitative reproducibility data for controlled, patterned strain in monolayer MoS$_2$ has also become available. Kumral \textit{et al.} \cite{kumral2025high} demonstrated spatially programmable biaxial strain up to 2.2\% via conformal transfer onto two-photon-lithography-patterned substrates, with 0.13\%~$\mu$m$^{-1}$ spatial resolution and stability maintained over months rather than decaying on device-relevant timescales, establishing that the fabrication route assumed in Sec.~\ref{sec:device} is realized with quantified spatial and temporal control at the $\sim$2\% level.

The $\varepsilon_{max} \approx 7.0\%$ peak strain required here exceeds this demonstrated value, and this paper does not claim it has been directly realized in a contact-integrated device. Comparable magnitudes have, however, been reached separately via sharp-tip nanoindentation, with localized elastic strains of 6.8--7\% measured beneath a sharp indenter tip in layered thin-film geometries \cite{zak2024controlling} and up to 2.6\% specifically in monolayer MoS$_2$ via AFM-based thermomechanical nanoindentation \cite{ThermoNano2023}. The $\sim$7\% strain assumed here therefore sits above what has been shown with simultaneous spatial programmability and stability, but within reach of sharp-tip techniques, and combining the two into a contact-integrated, lithographically defined device remains an open fabrication target rather than an established result, stated explicitly rather than implied as solved.

{Set against this fabrication gap, }the strain profile sensitivity analysis in Fig.~\ref{fig:sensitivity} demonstrates that the macroscopic VTMR signature is structurally robust, a property that relaxes the precision demanded of any specific piezoelectric actuator design. Deriving the exact strain field for a given PZT geometry, including its full piezoelectric coupling tensor, electrode architecture, and mechanical boundary conditions, is a structural mechanics problem this transport-focused framework deliberately treats as a downstream engineering task rather than a prerequisite. {The uniform and contact-peaked profiles yield 76.7\% and 63.6\% suppression respectively. While both remain large, the 13-percentage-point gap indicates that the exact profile shape within the channel interior has a measurable, though secondary, effect relative to the dominant contact contribution.} The critical physical requirement that the pseudo-magnetic field is concentrated at the metallic injection interfaces can be realized by any monotonically increasing strain distribution with a contact-peaked maximum, a condition achievable across a range of piezoelectric actuator geometries.

Beyond strain mechanics, the operational temperature of 77~K is a strict theoretical requirement for current experimental realization. At room temperature (300~K), thermal broadening and intense intervalley optical phonon scattering would severely degrade the unperturbed 3000~cm$^2$/Vs mobility \cite{Kaasbjerg2012}. This phonon dominance would obscure the pseudo-magnetic Lorentz deflections, suppressing the {63.6\%} VTMR signature into the background thermal noise. Consequently, while this framework proves that topological valley-sorting can be read electrically using standard measurement terminals, practical implementation remains restricted to cryogenic environments, consistent with the broader body of low-temperature strain- and field-based valleytronic characterization already established in the 2D-materials literature.

The transport simulations further assume an ultra-clean van der Waals interface reaching the intrinsic acoustic-phonon-limited mobility of 3000~cm$^2$/Vs at 77~K \cite{Kaasbjerg2012}. While early gated MoS$_2$ on SiO$_2$ was bottlenecked by oxide-trap scattering to cryogenic mobilities near 1000~cm$^2$/Vs \cite{Radisavljevic2013}, hBN encapsulation has since screened out this remote interfacial scattering, {with hBN-encapsulated heterostructures reported reaching cryogenic mobilities exceeding 34{,}000~cm$^2$/Vs for multilayers and} monolayer mobilities now routinely reaching the 3000~cm$^2$/Vs regime at cryogenic temperature \cite{Cui2015}, making the baseline mobility assumed here a realistic representation of state-of-the-art devices.

\subsection{Contact Transparency and the Thermionic-Emission Limit}
\label{sec:thermionic_limit}

Ultimately, the {63.6\%} transconductance suppression represents the intrinsic operational limit of the channel. In physical two-terminal measurements, the total device resistance is governed by the series equation $R_{tot} = 2R_c + R_{ch}(B_z)$, where $R_c$ is the interfacial contact resistance. Because of the Fermi-level pinning typically observed at transition metal and TMD interfaces, Schottky barriers can dominate the total resistance, effectively hiding the topological resistance variation ($\Delta R_{ch}$) modeled here.

 To quantify this constraint rather than treat it qualitatively, the barrier-limited contact resistance $R_B$ is evaluated directly from thermionic emission theory \cite{sze2008semiconductor} and compared against the independently verified $R_{c,V}$ and $\rho_s$ extracted in Sec.~\ref{sec:tlm}. For a metal-semiconductor Schottky barrier of height $\phi_B$, the injected current density under thermionic emission is:

\begin{equation}
J = A^{**}T^2 \exp\!\left(-\frac{q\phi_B}{k_BT}\right)\left[\exp\!\left(\frac{qV}{k_BT}\right)-1\right],
\label{eq:thermionic}
\end{equation}

where $A^{**} = A_0 (m^*/m_0)$ is the effective Richardson constant and $A_0 = 120$~A/(cm$^2$K$^2$) is the free-electron Richardson constant \cite{sze2008semiconductor}. Using $m^* = 0.45m_0$ as established in Sec.~\ref{sec:device}, $A^{**} = 54.08$~A/(cm$^2$K$^2$). Linearizing Eq.~\eqref{eq:thermionic} about $V=0$ gives the differential specific contact resistance,

\begin{equation}
\rho_c(\phi_B) = \left.\frac{\partial V}{\partial J}\right|_{V=0} = \frac{k_BT}{q A^{**}T^2}\exp\!\left(\frac{q\phi_B}{k_BT}\right).
\label{eq:rho_c}
\end{equation}

Converting $\rho_c$ to a per-unit-width series resistance using the 5~nm source/drain metal overlap length established in Sec.~\ref{sec:tlm} as the injection footprint,

\begin{equation}
R_B(\phi_B) = \frac{\rho_c(\phi_B)}{L_{contact}},
\label{eq:R_B}
\end{equation}

and the total two-terminal device resistance becomes,

\begin{equation}
R_{tot}(\phi_B) = \rho_s L + 2R_{c,V} + 2R_B(\phi_B),
\label{eq:R_tot_thermionic}
\end{equation}

using the verified $\rho_s = 5.944$~$\Omega$/sq and $R_{c,V} = 0.448$~$\Omega\cdot\mu$m from Sec.~\ref{sec:tlm} directly, rather than re-fitted values.

\begin{figure}[htbp]
\centering
\includegraphics[width=\linewidth]{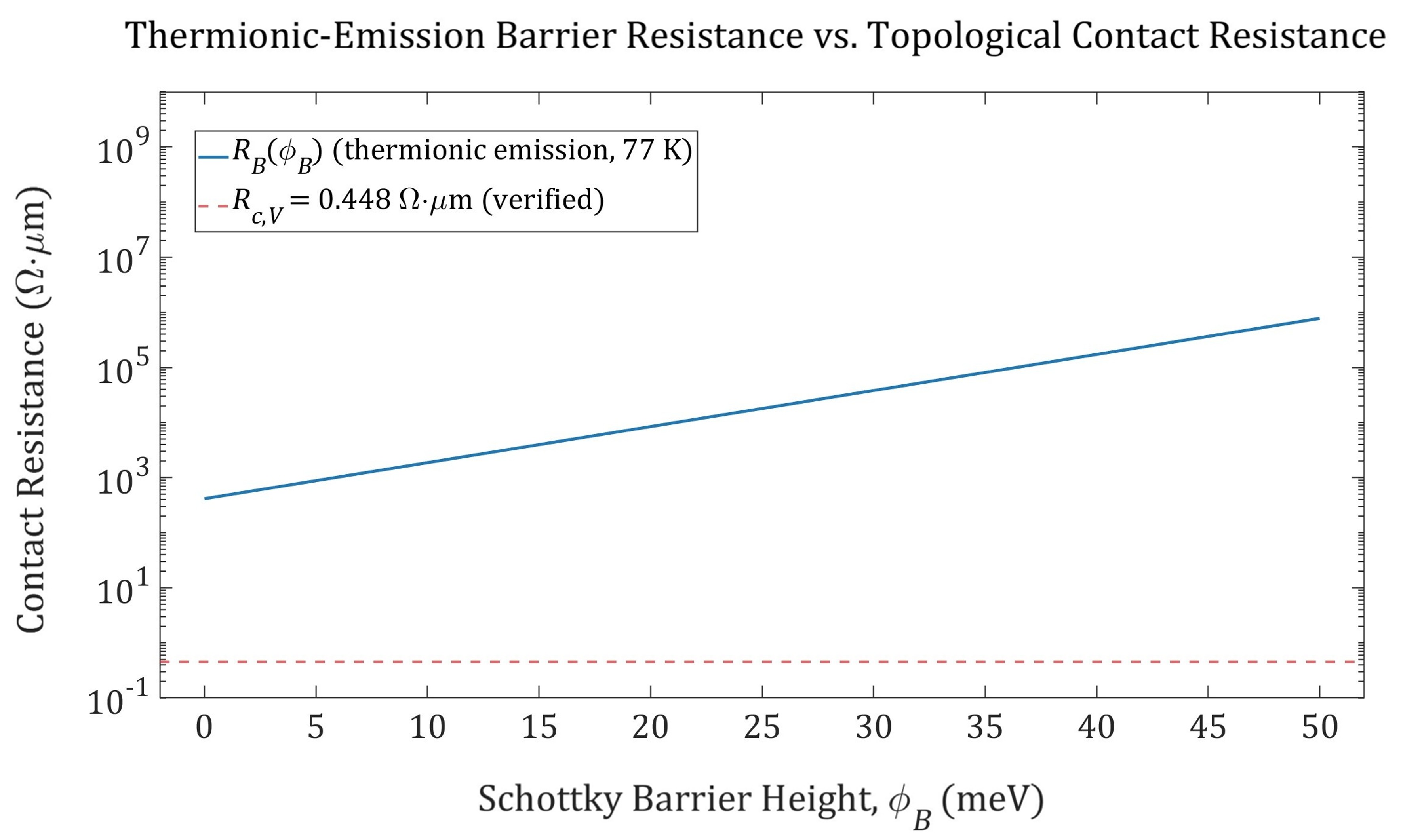}
\caption{Thermionic-emission contact resistance $R_B$ (blue solid line) evaluated as a function of the Schottky barrier height $\phi_B$ at 77~K, compared against the verified valley-topological contact resistance $R_{c,V}$ (red dashed line). Even at the idealized zero-barrier limit, thermionic emission limits the injected current to a regime where the topological signature is overwhelmed by almost three orders of magnitude.}
\label{fig:thermionic}
\end{figure}

At 77~K, $k_BT \approx 6.64$~meV, and evaluating Eq.~\eqref{eq:rho_c}-\eqref{eq:R_B} at $\phi_B = 0$ (the theoretical lower bound for a thermionic-emission contact) already gives $R_B(0) \approx 414$~$\Omega\cdot\mu$m, a factor of $\sim$924 larger than $R_{c,V}$ (Fig.~\ref{fig:thermionic}). The barrier resistance grows exponentially from this floor, reaching $R_B \approx 1.9\times10^3$~$\Omega\cdot\mu$m at $\phi_B = 10$~meV and $R_B \approx 7.8\times10^5$~$\Omega\cdot\mu$m at $\phi_B = 50$~meV, overwhelming $\rho_s L + 2R_{c,V}$ by many orders of magnitude across the entire physically relevant barrier range (Table~\ref{tab:thermionic}).

This shows that classical thermionic emission cannot support the current needed to observe VTMR at cryogenic temperature under any finite barrier height, including the ideal zero-barrier limit. Injection must instead proceed by field emission through a thin, tunneling-transparent depletion region, precisely the condition phase-engineered 1T-MoS$_2$ \cite{Kappera2014} and bismuth semi-metal contacts \cite{Shen2021} are designed to satisfy, making their citation here load-bearing rather than illustrative. This also bounds the $E_y=0$ idealization of Sec.~\ref{sec:theory}, since that condition requires tunneling-dominated injection rather than being assumed for convenience.

A complementary TCAD validation using an explicit workfunction-based Schottky boundary condition was also attempted. Reproducing the verified Ohmic baseline under that contact model requires solver-specific calibration beyond the scope of this work and is left for follow-on work. The analytical treatment above depends only on the independently verified $\rho_s$, $R_{c,V}$, and standard thermionic-emission theory, and is fully reproducible independent of that calibration.

\begin{table}[htbp]
\centering
\caption{Thermionic-emission contact resistance $R_B(\phi_B)$ at 77~K compared against the verified valley-topological contact resistance $R_{c,V} = 0.448$~$\Omega\cdot\mu$m. All resistances are per unit device width. \\}
\label{tab:thermionic}
\begin{tabular}{ccc}
\hline \hline
$\phi_B$ (meV) & $R_B$ ($\Omega\cdot\mu$m) & $R_B / R_{c,V}$ \\
\hline
0 & $4.14\times10^{2}$ & 924 \\
5 & $8.79\times10^{2}$ & 1963 \\
10 & $1.87\times10^{3}$ & 4170 \\
20 & $8.43\times10^{3}$ & 18821 \\
30 & $3.81\times10^{4}$ & 84950 \\
50 & $7.75\times10^{5}$ & $1.73\times10^{6}$ \\
100 & $1.45\times10^{9}$ & $3.24\times10^{9}$ \\
\hline \hline
\end{tabular}
\end{table}

However, experimental verification of this macroscopic VTMR is now practically achievable. Advanced contact engineering, such as phase-engineered 1T-MoS$_2$ metallic boundaries \cite{Kappera2014} or bismuth semi-metal electrodes \cite{Shen2021}, has proven to significantly suppress Fermi-level pinning and yield highly transparent Ohmic contacts that mitigate barrier effects even at reduced temperatures. These state-of-the-art injection mechanisms are the clearest available path to satisfying the requirement $R_{ch} \gg 2R_c$ as contact engineering for 2D materials continues to mature. This ensures the transport bottleneck remains entirely within the topologically strained active region, providing a clean macroscopic readout of the underlying quantum valley mechanics.

\subsection{Distinguishing the Valley-Topological Signature from Competing Transport Mechanisms}

Because several independent mechanisms, most notably strain-induced bandgap modulation and contact alloying or degradation, can also alter the ON-state current of a strained device, isolating the valley-topological contribution from these competing effects requires more than observing a current suppression alone. Four independent, experimentally accessible signatures distinguish VTMR from these alternative origins.

\textbf{Subthreshold invariance.} A bandgap shift would be expected to alter the threshold voltage and subthreshold swing, both set by band alignment at the channel-oxide interface. Fig.~\ref{fig:highk}(a) shows the strained and unstrained devices share a common threshold ($\sim$1.0~V) and subthreshold swing ($\sim$102~meV/dec): the suppression is confined to the ON-state transconductance, inconsistent with a bandgap-shift origin.

\textbf{Temperature dependence.} Because the suppression scales with $\mu_0 B_z$ (Eq.~\eqref{eq:mobility_tensor}) and $\mu_0$ strengthens as $T$ decreases, VTMR is predicted to increase with further cooling, tracking $\mu_0(T)$, the opposite trend expected from contact alloying, which worsens as thermionic/defect-assisted conduction freezes out. A temperature sweep of $R_{tot}(L,T)$ on the same TLM structures (Sec.~\ref{sec:tlm}) directly discriminates the two.

\textbf{Gate-voltage scaling.} The same strain profile gives 34.1\% suppression Ohmic (Sec.~\ref{sec:ohmic}) versus 63.6\% gated (Sec.~\ref{sec:gated}), an amplification attributed in Sec.~\ref{sec:gated} to gate-controlled injection velocity, though this attribution is not independently decomposed and gate-field-dependent Schottky barrier lowering could plausibly produce a similar Ohmic-to-gated amplification, so the signature is suggestive rather than conclusive pending the dedicated measurement proposed above.

\textbf{Nonlocal transverse measurement.} The valley-Hall deflection underlying this mechanism (Sec.~\ref{sec:theory}) separately predicts a nonlocal transverse voltage of the type established for valley-Hall transport \cite{Mak2014, Gorbachev2014}. Bandgap modulation and contact alloying are both scalar effects and predict no such signal, so a combined local/nonlocal measurement on the same device is a fully independent cross-check.

Together, these four tests, subthreshold invariance, temperature scaling, gate-regime amplification, and nonlocal cross-validation, are all obtainable from the same multi-length TLM device geometry already proposed for the primary $R_{c,V}$ extraction (Sec.~\ref{sec:tlm}), and require no additional device architectures beyond those already validated in this work.

\section{Conclusion}

Controlling and reading out valley-selective transport is a central goal of the wider 2D-materials community, yet whether this quantum-mechanical control survives inside a real, fully contacted device has remained largely inaccessible to existing modeling approaches. This work carries strain-induced quantum valley deflection out of the isolated nanoscale simulations where it has previously lived and into a full, contact-inclusive device prediction. By mapping the strain-induced vector potential of the Dirac Hamiltonian onto a localized phenomenological mobility tensor, a pseudo-magnetic scattering mechanism was successfully integrated into a commercial finite-element Poisson solver, bypassing the prohibitive computational cost of full non-equilibrium phase calculations and allowing this class of quantum valley effect to be predicted, for the first time, at the scale of a complete gated device with realistic metal contacts.

The macroscopic transport response under these topological conditions is characterized by simulations of a 300~nm MoS$_2$ field-effect transistor at 77~K. The channel is driven into deep inversion by the integration of a high-$\kappa$ gate dielectric, yielding a {63.6\%} suppression in the linear-regime ON-state current at a {6.14~T} gauge limit. A clear macroscopic readout for valley polarization is established by this effect, while the baseline subthreshold switching of the transistor is fully preserved.

Alongside this, the exact physical origin of this suppression is resolved by a targeted strain profile sensitivity analysis. The macroscopic transconductance drop is structurally governed by the localized carrier injection interfaces rather than the bulk transport channel. The full {63.6\%} suppression is reproduced when the strain is modulated exclusively at the metallic boundaries, whereas the behavior is reduced to $<$1\% when the gauge field is confined entirely to the channel interior. Subject to the theoretical constraints of cryogenic operation and highly transparent Ohmic interfaces, it is proven by this contact-specific behavior that valley polarization can be isolated and measured using steady-state longitudinal currents strictly by engineering the injection boundaries. 

This contact-localized origin is quantitatively formalized through a transmission-line-method extraction of the channel-length dependence, which isolates a single geometry-independent valley-topological contact resistance, $R_{c,V} = 0.448~\Omega\cdot\mu\text{m}$, and defines a predictive crossover length $L^\ast \approx 151$~nm governing the scaling of the observable VTMR signal with device geometry. This scaling law offers a concrete and testable design principle. Shorter-channel Ohmic test structures more efficiently isolate the topological contact signature from dilution by the unperturbed bulk channel, giving experimentalists a direct lever for maximizing signal visibility independent of the underlying strain mechanics.

Ultimately, a computationally scalable pathway is provided by this framework for embedding quantum valley-deflection physics into standard device-scale simulation environments, enabling full-device-scale prediction of strain-engineered valleytronic behavior without the use of prohibitive non-equilibrium quantum transport calculations. Furthermore, a specific and actionable experimental target for the broader 2D-materials community working toward strain-engineered valleytronic devices is offered by the identification of the metal-semiconductor contact interface as the governing structural element.

\section*{Data Availability Statement}

The complete multi-physics computational framework, including the custom finite-element tensor scripts and the structural mechanics datasets generated during this study, are openly available in the Zenodo repository at \href{https://doi.org/10.5281/zenodo.21346193}{https://doi.org/10.5281/zenodo.21346193}.

\begin{acknowledgments}
The author acknowledges the computational resources and simulation frameworks provided by the Department of Electrical and Electronics Engineering at BITS Pilani, Dubai Campus. 
\end{acknowledgments}

\bibliographystyle{apsrev4-1}
\bibliography{ref}

@article{Theis2010,
  title     = {It's Time to Reinvent the Transistor!},
  author    = {Theis, Thomas N. and Solomon, Paul M.},
  journal   = {Science},
  volume    = {327},
  number    = {5973},
  pages     = {1600--1601},
  year      = {2010},
  doi       = {10.1126/science.1187597},
  publisher = {American Association for the Advancement of Science}
}

@article{Nikonov2013,
  title     = {Overview of Beyond-{CMOS} Devices and a Uniform Methodology for Their Benchmarking},
  author    = {Nikonov, Dmitri E. and Young, Ian A.},
  journal   = {Proceedings of the IEEE},
  volume    = {101},
  number    = {12},
  pages     = {2498--2533},
  year      = {2013},
  doi       = {10.1109/JPROC.2013.2268395},
  publisher = {IEEE}
}

@article{Rycerz2007,
  title     = {Valley filter and valley valve in graphene},
  author    = {Rycerz, A. and Tworzyd{\l}o, J. and Beenakker, C. W. J.},
  journal   = {Nature Physics},
  volume    = {3},
  number    = {3},
  pages     = {172--175},
  year      = {2007},
  doi       = {10.1038/nphys547},
  publisher = {Nature Publishing Group}
}

@book{sze2008semiconductor,
  title={Semiconductor devices: physics and technology},
  author={Sze, Simon Min},
  year={2008},
  publisher={John wiley \& sons}
}

@article{Schaibley2016,
  title     = {Valleytronics in {2D} materials},
  author    = {Schaibley, John R. and Yu, Hongyi and Clark, Genevieve
               and Rivera, Pasqual and Ross, Jason S. and Seyler, Kyle L.
               and Yao, Wang and Xu, Xiaodong},
  journal   = {Nature Reviews Materials},
  volume    = {1},
  number    = {11},
  pages     = {16055},
  year      = {2016},
  doi       = {10.1038/natrevmats.2016.55},
  publisher = {Nature Publishing Group}
}

@article{Xiao2012,
  title     = {Coupled Spin and Valley Physics in Monolayers of {MoS}$_2$
               and Other Group-{VI} Dichalcogenides},
  author    = {Xiao, Di and Liu, Gui-Bin and Feng, Wanxiang
               and Xu, Xiaodong and Yao, Wang},
  journal   = {Physical Review Letters},
  volume    = {108},
  number    = {19},
  pages     = {196802},
  year      = {2012},
  doi       = {10.1103/PhysRevLett.108.196802},
  publisher = {American Physical Society}
}

@article{Mak2014,
  title     = {The valley {H}all effect in {MoS}$_2$ transistors},
  author    = {Mak, Kin Fai and McGill, Kathryn L. and Park, Jiwoong
               and McEuen, Paul L.},
  journal   = {Science},
  volume    = {344},
  number    = {6191},
  pages     = {1489--1492},
  year      = {2014},
  doi       = {10.1126/science.1250140},
  publisher = {American Association for the Advancement of Science}
}

@article{Cao2012,
  title     = {Valley-selective circular dichroism of monolayer
               molybdenum disulphide},
  author    = {Cao, Ting and Wang, Gang and Han, Wenpeng and Ye, Huiqiao
               and Zhu, Chuanrui and Shi, Jianrong and Niu, Qian
               and Tan, Pingheng and Wang, Enge and Liu, Baoli
               and Feng, Ji},
  journal   = {Nature Communications},
  volume    = {3},
  pages     = {887},
  year      = {2012},
  doi       = {10.1038/ncomms1882},
  publisher = {Nature Publishing Group}
}

@article{Zeng2012,
  title     = {Valley polarization in {MoS}$_2$ monolayers by optical pumping},
  author    = {Zeng, Hualing and Dai, Junfeng and Yao, Wang
               and Xiao, Di and Cui, Xiaodong},
  journal   = {Nature Nanotechnology},
  volume    = {7},
  number    = {8},
  pages     = {490--493},
  year      = {2012},
  doi       = {10.1038/nnano.2012.95},
  publisher = {Nature Publishing Group}
}

@article{Guinea2010,
  title     = {Energy gaps and a zero-field quantum {H}all effect in
               graphene by strain engineering},
  author    = {Guinea, F. and Katsnelson, M. I. and Geim, A. K.},
  journal   = {Nature Physics},
  volume    = {6},
  number    = {1},
  pages     = {30--33},
  year      = {2010},
  doi       = {10.1038/nphys1420},
  publisher = {Nature Publishing Group}
}

@article{Vozmediano2010,
  title     = {Gauge fields in graphene},
  author    = {Vozmediano, M. A. H. and Katsnelson, M. I. and Guinea, F.},
  journal   = {Physics Reports},
  volume    = {496},
  number    = {4-5},
  pages     = {109--148},
  year      = {2010},
  doi       = {10.1016/j.physrep.2010.07.003},
  publisher = {Elsevier}
}

@article{Settnes2016,
  title     = {Pseudomagnetic fields and triaxial strain in graphene},
  author    = {Settnes, Mikkel and Power, Stephen R. and Jauho, Antti-Pekka},
  journal   = {Physical Review B},
  volume    = {93},
  number    = {3},
  pages     = {035456},
  year      = {2016},
  doi       = {10.1103/PhysRevB.93.035456},
  publisher = {American Physical Society}
}

@article{Levy2010,
  title     = {Strain-Induced Pseudo--Magnetic Fields Greater Than 300 {T}esla
               in Graphene Nanobubbles},
  author    = {Levy, N. and Burke, S. A. and Meaker, K. L.
               and Panlasigui, M. and Zettl, A. and Guinea, F.
               and {Castro Neto}, A. H. and Crommie, M. F.},
  journal   = {Science},
  volume    = {329},
  number    = {5991},
  pages     = {544--547},
  year      = {2010},
  doi       = {10.1126/science.1191700},
  publisher = {American Association for the Advancement of Science}
}

@article{Low2014,
  title     = {Strain-Induced Pseudomagnetic Field for Novel
               Graphene Electronics},
  author    = {Low, Tony and Guinea, Francisco},
  journal   = {Nano Letters},
  volume    = {10},
  number    = {9},
  pages     = {3551--3554},
  year      = {2010},
  doi       = {10.1021/nl1018063},
  publisher = {ACS Publications}
}

@article{Beconcini2016,
  title     = {Nonlocal topological valley transport at large valley
               {H}all angles},
  author    = {Beconcini, Michael and Polini, Marco and Taddei, Fabio},
  journal   = {Physical Review B},
  volume    = {94},
  number    = {11},
  pages     = {115441},
  year      = {2016},
  doi       = {10.1103/PhysRevB.94.115441},
  publisher = {American Physical Society}
}

@article{Gorbachev2014,
  title     = {Detecting topological currents in graphene superlattices},
  author    = {Gorbachev, R. V. and Song, J. C. W. and Yu, G. L.
               and Kretinin, A. V. and Withers, F. and Cao, Y.
               and Mishchenko, A. and Grigorieva, I. V.
               and Novoselov, K. S. and Levitov, L. S. and Geim, A. K.},
  journal   = {Science},
  volume    = {346},
  number    = {6208},
  pages     = {448--451},
  year      = {2014},
  doi       = {10.1126/science.1254966},
  publisher = {American Association for the Advancement of Science}
}

@article{Cazalilla2014,
  title     = {Quantum Spin {H}all Effect in Two-Dimensional Crystals
               of Transition-Metal Dichalcogenides},
  author    = {Cazalilla, M. A. and Ochoa, H. and Guinea, F.},
  journal   = {Physical Review Letters},
  volume    = {113},
  number    = {7},
  pages     = {077201},
  year      = {2014},
  doi       = {10.1103/PhysRevLett.113.077201},
  publisher = {American Physical Society}
}

@article{CarrilloBastos2014,
  title = {Strained fold-assisted transport in graphene systems},
  author = {Carrillo-Bastos, R. and Le\'on, C. and Faria, D. and Latg\'e, A. and Andrei, E. Y. and Sandler, N.},
  journal = {Phys. Rev. B},
  volume = {94},
  issue = {12},
  pages = {125422},
  numpages = {6},
  year = {2016},
  month = {Sep},
  publisher = {American Physical Society},
  doi = {10.1103/PhysRevB.94.125422},
  url = {https://link.aps.org/doi/10.1103/PhysRevB.94.125422}
}

@article{Rostami2015,
  title     = {Theory of strain in single-layer transition metal
               dichalcogenides},
  author    = {Rostami, Habib and Rold\'an, Rafael and Cappelluti, Emmanuele
               and Asgari, Reza and Guinea, Francisco},
  journal   = {Physical Review B},
  volume    = {92},
  number    = {19},
  pages     = {195402},
  year      = {2015},
  doi       = {10.1103/PhysRevB.92.195402},
  publisher = {American Physical Society}
}

@article{dean2010boron,
  title={Boron nitride substrates for high-quality graphene electronics},
  author={Dean, Cory R and Young, Andrea F and Meric, Inanc and Lee, Chris and Wang, Lei and Sorgenfrei, Sebastian and Watanabe, Kenji and Taniguchi, Takashi and Kim, Phillip and Shepard, Kenneth L and others},
  journal={Nature nanotechnology},
  volume={5},
  number={10},
  pages={722--726},
  year={2010},
  doi={https://doi.org/10.1038/nnano.2010.172},
  publisher={Nature Publishing Group}
}

@article{ando1982electronic,
  title={Electronic properties of two-dimensional systems},
  author={Ando, Tsuneya and Fowler, Alan B and Stern, Frank},
  journal={Reviews of Modern Physics},
  volume={54},
  number={2},
  pages={437},
  year={1982},
  doi={https://doi.org/10.1103/RevModPhys.54.437},
  publisher={APS}
}

@article{Scharfetter1969,
  title     = {Large-signal analysis of a silicon {R}ead diode oscillator},
  author    = {Scharfetter, Donald L. and Gummel, Hermann K.},
  journal   = {IEEE Transactions on Electron Devices},
  volume    = {16},
  number    = {1},
  pages     = {64--77},
  year      = {1969},
  doi       = {10.1109/T-ED.1969.16566},
  publisher = {IEEE}
}

@article{Amorim2016,
  title     = {Novel effects of strains in graphene and other two
               dimensional materials},
  author    = {Amorim, B. and Cortijo, A. and {de Juan}, F.
               and Grushin, A. G. and Guinea, F.
               and Guti{\'e}rrez-Rubio, A. and Ochoa, H.
               and Parente, V. and Rold\'an, R. and {San-Jos\'e}, P.
               and Schiefele, J. and Sturla, M.
               and Vozmediano, M. A. H.},
  journal   = {Physics Reports},
  volume    = {617},
  pages     = {1--54},
  year      = {2016},
  doi       = {10.1016/j.physrep.2015.12.006},
  publisher = {Elsevier}
}

@book{Lundstrom2000,
  title     = {Fundamentals of Carrier Transport},
  author    = {Lundstrom, Mark},
  year      = {2000},
  edition   = {2nd},
  publisher = {Cambridge University Press},
  address   = {Cambridge}
}

@article{Kaasbjerg2012,
  title = {Acoustic phonon limited mobility in two-dimensional semiconductors: Deformation potential and piezoelectric scattering in monolayer MoS${}_{2}$ from first principles},
  author = {Kaasbjerg, Kristen and Thygesen, Kristian S. and Jauho, Antti-Pekka},
  journal = {Phys. Rev. B},
  volume = {87},
  issue = {23},
  pages = {235312},
  numpages = {15},
  year = {2013},
  month = {Jun},
  publisher = {American Physical Society},
  doi = {10.1103/PhysRevB.87.235312},
  url = {https://link.aps.org/doi/10.1103/PhysRevB.87.235312}
}

@article{Bertolazzi2011,
  title     = {Stretching and Breaking of Ultrathin {MoS}$_2$},
  author    = {Bertolazzi, Simone and Brivio, Jacopo and Kis, Andras},
  journal   = {ACS Nano},
  volume    = {5},
  number    = {12},
  pages     = {9703--9709},
  year      = {2011},
  doi       = {10.1021/nn203879f},
  publisher = {ACS Publications}
}

@article{Radisavljevic2011,
  title     = {Single-layer {MoS}$_2$ transistors},
  author    = {Radisavljevic, B. and Radenovic, A. and Brivio, J.
               and Giacometti, V. and Kis, A.},
  journal   = {Nature Nanotechnology},
  volume    = {6},
  number    = {3},
  pages     = {147--150},
  year      = {2011},
  doi       = {10.1038/nnano.2010.279},
  publisher = {Nature Publishing Group}
}

@book{Taur2013,
  title     = {Fundamentals of Modern {VLSI} Devices},
  author    = {Taur, Yuan and Ning, Tak H.},
  year      = {2013},
  edition   = {2nd},
  publisher = {Cambridge University Press},
  address   = {Cambridge}
}

@article{Peelaers2012,
  title     = {Effects of strain on band structure and effective masses
               in {MoS}$_2$},
  author    = {Peelaers, Hartwin and {Van de Walle}, Chris G.},
  journal   = {Physical Review B},
  volume    = {86},
  number    = {24},
  pages     = {241401},
  year      = {2012},
  doi       = {10.1103/PhysRevB.86.241401},
  publisher = {American Physical Society}
}

@article{Radisavljevic2013,
  title     = {Mobility engineering and a metal--insulator transition
               in monolayer {MoS}$_2$},
  author    = {Radisavljevic, Branimir and Kis, Andras},
  journal   = {Nature Materials},
  volume    = {12},
  number    = {9},
  pages     = {815--820},
  year      = {2013},
  doi       = {10.1038/nmat3687},
  publisher = {Nature Publishing Group}
}

@article{Jiang2017,
  title     = {Visualizing Strain-Induced Pseudomagnetic Fields in
               Graphene through an {hBN} Magnifying Glass},
  author    = {Jiang, Yuhang and Mao, Jinhai and Duan, Junxi
               and Lai, Xinyuan and Watanabe, Kenji and Taniguchi, Takashi
               and Andrei, Eva Y.},
  journal   = {Nano Letters},
  volume    = {17},
  number    = {5},
  pages     = {2839--2843},
  year      = {2017},
  doi       = {10.1021/acs.nanolett.6b05228},
  publisher = {ACS Publications}
}

@article{Kappera2014,
  title     = {Phase-engineered low-resistance contacts for ultrathin
               {MoS}$_2$ transistors},
  author    = {Kappera, Rajesh and Voiry, Damien and Yalcin, Sibel Ebru
               and Branch, Brittany and Gupta, Gautam
               and Mohite, Aditya D. and Chhowalla, Manish},
  journal   = {Nature Materials},
  volume    = {13},
  number    = {12},
  pages     = {1128--1134},
  year      = {2014},
  doi       = {10.1038/nmat4080},
  publisher = {Nature Publishing Group}
}

@article{Shen2021,
  title     = {Ultralow contact resistance between semimetal and
               monolayer semiconductors},
  author    = {Shen, Pin-Chun and Su, Cong and Lin, Yuxuan
               and Chou, Ang-Sheng and Cheng, Chao-Ching
               and Park, Ji-Hoon and Chiu, Ming-Hui and Lu, Ang-Yu
               and Tang, Hao-Ling and Tavakoli, Mohammad Mahdi
               and Pitner, Gregory and Ji, Xiang and Cai, Zhengyang
               and Mao, Nannan and Wang, Jiangtao and Tung, Vincent
               and Li, Ju and Bokor, Jeffrey and Zettl, Alex
               and Wu, Chih-I and Palacios, Tom\'as
               and Li, Lain-Jong and Kong, Jing},
  journal   = {Nature},
  volume    = {593},
  number    = {7858},
  pages     = {211--217},
  year      = {2021},
  doi       = {10.1038/s41586-021-03472-9},
  publisher = {Nature Publishing Group}
}

@book{Ashcroft1976,
  title     = {Solid State Physics},
  author    = {Ashcroft, Neil W. and Mermin, N. David},
  year      = {1976},
  publisher = {Holt, Rinehart and Winston},
  address   = {New York}
}

@article{Ghatak2011,
  title     = {Nature of Electronic States in Atomically Thin
               {MoS}$_2$ Field-Effect Transistors},
  author    = {Ghatak, Subhamoy and Pal, Aswini N. and Ghosh, Arindam},
  journal   = {ACS Nano},
  volume    = {5},
  number    = {10},
  pages     = {7707--7712},
  year      = {2011},
  doi       = {10.1021/nn202852j},
  publisher = {ACS Publications}
}

@book{Selberherr1984,
  title     = {Analysis and Simulation of Semiconductor Devices},
  author    = {Selberherr, Siegfried},
  year      = {1984},
  publisher = {Springer-Verlag},
  address   = {Vienna}
}

@article{Conley2013,
  author    = {Conley, Hiram J. and Wang, Bin and Ziegler, Jed I.
               and Haglund, Richard F. and Pantelides, Sokrates T.
               and Bolotin, Kirill I.},
  title     = {Bandgap Engineering of Strained Monolayer and Bilayer
               {MoS}$_2$},
  journal   = {Nano Letters},
  year      = {2013},
  volume    = {13},
  number    = {8},
  pages     = {3626--3630},
  doi       = {10.1021/nl4014748},
  publisher = {ACS Publications}
}

@article{Cresti2016,
  author    = {Cresti, Alessandro and Nikoli{\'c}, Branislav K.
               and Garc{\'\i}a, Jose Hugo and Roche, Stephan},
  title     = {Charge, Spin and Valley {H}all Effects in Disordered
               Graphene},
  journal   = {La Rivista del Nuovo Cimento},
  year      = {2016},
  volume    = {39},
  number    = {12},
  pages     = {587--667},
  doi       = {10.1393/ncr/i2016-10130-6},
  publisher = {Societ{\`a} Italiana di Fisica}
}

@article{Cui2015,
  title     = {Multi-terminal transport measurements of {MoS}$_2$ using a
               van der {Waals} heterostructure device platform},
  author    = {Cui, Xu and Lee, Gwan-Hyoung and Kim, Young Duck
               and Arefe, Ghidewon and Huang, Pinshane Y.
               and Lee, Chul-Ho and Chenet, Daniel A. and Zhang, Xian
               and Wang, Lei and Ye, Fan and Hone, James},
  journal   = {Nature Nanotechnology},
  volume    = {10},
  number    = {6},
  pages     = {534--540},
  year      = {2015},
  doi       = {10.1038/nnano.2015.70},
  publisher = {Nature Publishing Group}
}

@article{Yang2021,
  title     = {Strain engineering of two-dimensional materials: Methods, properties, and applications},
  author    = {Yang, Shengxue and Chen, Yujia and Jiang, Chengbao},
  journal   = {InfoMat},
  volume    = {3},
  number    = {4},
  pages     = {397--420},
  year      = {2021},
  doi       = {10.1002/inf2.12177},
  publisher = {Wiley}
}

@article{lloyd2016band,
  title={Band gap engineering with ultralarge biaxial strains in suspended monolayer MoS2},
  author={Lloyd, David and Liu, Xinghui and Christopher, Jason W and Cantley, Lauren and Wadehra, Anubhav and Kim, Brian L and Goldberg, Bennett B and Swan, Anna K and Bunch, J Scott},
  journal={Nano letters},
  volume={16},
  number={9},
  pages={5836--5841},
  year={2016},
  doi={10.1021/acs.nanolett.6b02615},
  publisher={ACS Publications}
}

@article{schmidt2016reversible,
  title={Reversible uniaxial strain tuning in atomically thin WSe2},
  author={Schmidt, Robert and Niehues, Iris and Schneider, Robert and Drueppel, Matthias and Deilmann, Thorsten and Rohlfing, Michael and De Vasconcellos, Steffen Michaelis and Castellanos-Gomez, Andres and Bratschitsch, Rudolf},
  journal={2D Materials},
  volume={3},
  number={2},
  pages={021011},
  year={2016},
  doi={10.1088/2053-1583/3/2/021011},
  publisher={IOP Publishing}
}

@article{castellanos2013local,
  title={Local strain engineering in atomically thin MoS2},
  author={Castellanos-Gomez, Andres and Rold{\'a}n, Rafael and Cappelluti, Emmanuele and Buscema, Michele and Guinea, Francisco and Van Der Zant, Herre SJ and Steele, Gary A},
  journal={Nano letters},
  volume={13},
  number={11},
  pages={5361--5366},
  year={2013},
  doi={https://doi.org/10.1021/nl402875m},
  publisher={ACS Publications}
}

@article{Xiao2010,
  title = {Berry phase effects on electronic properties},
  author = {Xiao, Di and Chang, Ming-Che and Niu, Qian},
  journal = {Rev. Mod. Phys.},
  volume = {82},
  issue = {3},
  pages = {1959--2007},
  year = {2010},
  publisher = {American Physical Society}
}

@article{Shockley1952,
  title = {Statistics of the Recombinations of Holes and Electrons},
  author = {Shockley, W. and Read, W. T.},
  journal = {Phys. Rev.},
  volume = {87},
  issue = {5},
  pages = {835--842},
  year = {1952},
  publisher = {American Physical Society}
}

@article{Hall1952,
  title = {Electron-Hole Recombination in Germanium},
  author = {Hall, R. N.},
  journal = {Phys. Rev.},
  volume = {87},
  issue = {2},
  pages = {387--387},
  year = {1952},
  publisher = {American Physical Society}
}

@article{Berger1972,
  author  = {Berger, H. H.},
  title   = {Models for contacts to planar devices},
  journal = {Solid-State Electronics},
  volume  = {15},
  number  = {2},
  pages   = {145--158},
  year    = {1972},
  doi     = {10.1016/0038-1101(72)90048-2}
}

@article{ReevesHarrison1982,
  author  = {Reeves, G. K. and Harrison, H. B.},
  title   = {Obtaining the specific contact resistance from transmission line model measurements},
  journal = {IEEE Electron Device Letters},
  volume  = {3},
  pages   = {111--113},
  year    = {1982},
  doi     = {10.1109/EDL.1982.25502}
}

@article{Sekine2018,
  author  = {Sekine, Akihiko and MacDonald, Allan H.},
  title   = {Valley-dependent magnetoresistance in two-dimensional semiconductors},
  journal = {Physical Review B},
  volume  = {97},
  pages   = {201301},
  year    = {2018},
  doi     = {10.1103/PhysRevB.97.201301}
}

@article{kumral2025high,
  title={High-magnitude, spatially programmable, and sustained strain engineering of 2D semiconductors},
  author={Kumral, Boran and Demingos, Pedro Guerra and Serles, Peter and Yang, Shuo and Kim, Da Bin and Yu, Dian and Nair, Akhil and Rastogi, Akshat and Barri, Nima and Islam, Md Akibul and others},
  journal={arXiv preprint arXiv:2508.00972},
  year={2025}
}

@article{zak2024controlling,
  title={Controlling strain localization in thin films with nanoindenter tip sharpness},
  author={Zak, Stanislav},
  journal={Scientific reports},
  volume={14},
  number={1},
  pages={25500},
  year={2024},
  publisher={Nature Publishing Group UK London}
}

@article{ThermoNano2023,
    author = {Chang, Shunyu and Yan, Yongda and Geng, Yanquan},
    title = {Local Nanostrain Engineering of Monolayer MoS2 Using Atomic Force Microscopy-Based Thermomechanical Nanoindentation},
    journal = {Nano Letters},
    volume = {23},
    number = {20},
    pages = {9219-9226},
    year = {2023},
    month = {10},
    issn = {1530-6984},
    doi = {10.1021/acs.nanolett.3c01809},
    url = {https://doi.org/10.1021/acs.nanolett.3c01809},
    eprint = {https://pubs.acs.org/nalefd/article-pdf/23/20/9219/2646874/nl3c01809.pdf},
}

@misc{mos2emc2026,
      title={Multiscale ensemble Monte Carlo of transport and gas sensing in monolayer MoS2},
      author={Lado Filipovic},
      year={2026},
      eprint={2607.18895},
      archivePrefix={arXiv},
      primaryClass={cond-mat.mes-hall},
      url={https://arxiv.org/abs/2607.18895}
}

\end{document}